\documentclass[twocolumn,floatfix,showpacs,prd,aps,eqsecnum,tightenlines]{revtex4}
\usepackage{graphicx}
\usepackage{dcolumn}
\usepackage{bm}
\usepackage{hyperref}
\usepackage{amssymb}

\newcommand {\be}{\begin{equation}}
\newcommand {\ee}{\end{equation}}
\newcommand {\ba}{\begin{array}}
\newcommand {\ea}{\end{array}}
\newcommand {\bea}{\begin{eqnarray}}
\newcommand {\eea}{\end{eqnarray}}
\newcommand {\bean}{\begin{eqnarray*}}
\newcommand {\eean}{\end{eqnarray*}}

\newcommand {\vecnorm}[1]{\parallel \vec{#1} \parallel}

\newcommand {\sth}{\sin\theta}
\newcommand {\ssth}{\sin^2\theta}
\newcommand {\cth}{\cos\theta}
\newcommand {\ccth}{\cos^2\theta}
\newcommand {\sph}{\sin\phi}
\newcommand {\ssph}{\sin^2\phi}
\newcommand {\cph}{\cos\phi}
\newcommand {\ccph}{\cos^2\phi}

\begin{document}


\title[working title]{The Lazarus Project. II. Space-like extraction with
the quasi-Kinnersley tetrad}

\author{Manuela Campanelli}
\author{Bernard Kelly}
\author{Carlos O. Lousto}
\affiliation{Department of Physics and Astronomy,
and Center for Gravitational Wave Astronomy,
The University of Texas at Brownsville, 80 Fort Brown, Brownsville, TX 78520, USA}

\date{\today}

\begin{abstract}
The Lazarus project was designed to make the most of limited 3D binary
black-hole simulations, through the identification of perturbations at
late times, and subsequent evolution of the Weyl scalar $\Psi_4$ via
the Teukolsky formulation. Here we report on new developments,
employing the concept of the ``quasi-Kinnersley'' (transverse) frame,
valid in the full nonlinear regime, to analyze late-time numerical
space-times that should differ only slightly from Kerr. This allows us
to extract the essential information about the background Kerr
solution, and through this, to identify the radiation present. We
explicitly test this procedure with full-numerical evolutions of
Bowen-York data for single spinning black holes, head-on and orbiting
black holes near the ISCO regime. {These techniques can be
compared with previous Lazarus results, providing a measure of the
numerical-tetrad errors intrinsic to the method, and giving as a
by-product a more robust wave extraction method for numerical
relativity.}

\end{abstract}

\maketitle

\section{Introduction}\label{Sec:Intro}


The strong-field interaction of black-hole binary systems --- from
early approach through capture, mutual orbit and eventual merger, to
ring-down of the end-state single hole --- is expected to be a primary
source of gravitational radiation at all frequency scales, and has
been a focus of theoretical and numerical attention for forty
years. Early perturbative studies \cite{Price94a,Pullin98a,Lousto99a}
and two-dimensional numerical evolutions of axisymmetric binaries
(head-on collisions) \cite{Hahn64,Smarr75,Smarr76,Cadez71,Cadez74,Anninos98a}
were successful in producing late-stage waveforms representing
gravitational radiation. However, the move to full 3D simulations of
more general initial-data sets has proved extremely
difficult. Evolutions using the ``standard ADM'' 3+1 decomposition of
Einstein's equations and simple zero-shift gauge conditions have
stable lifetimes of $\sim 15-30 M$ (where $M$ is the total mass of the
space-time), far too short a time to complete a useful physical
simulations, much less extract the gravitational radiation emitted ---
the ultimate aim of numerical source simulations.


The Lazarus project \cite{Baker00b,Baker:2001sf} was conceived in the
context of such limitations. Working under the assumption that the
late-evolution 3+1 data can be considered a perturbation of a single
Kerr black hole, Lazarus extracts the radiation content everywhere in
the numerical domain, and uses it as initial data for a Teukolsky
perturbative evolution. In this manner, the original simulation may be
extended almost indefinitely, long enough to capture the entire
development of the outgoing radiation.

The beginning of the full-numerical simulation can also be interfaced
with a {\it far limit} approximation method and similar techniques to
evaluate a common regime of applicability can be developed
\cite{Baker:2001sf}.  Here, for the sake of definiteness, we will
assume a set of initial data as providing this interface values and
focus on the full-numerical / close limit matching.


Lazarus has been very successful, producing the first convergent wave
forms
\cite{Baker00b,Baker:2001nu,Baker:2002qf,Baker:2004wv,Campanelli:2004zw}
from 3D evolutions. In most cases, a ``plateau'' was identified --- a
range of extraction times $T$ where the emitted energy remained flat
and consistent. This plateau begins when the 3+1 data is linearly
perturbed from Kerr, and should end only when the radiation has begun
to leave the 3+1 numerical domain altogether; at this latter time,
Teukolsky extraction will no longer capture the full radiation
content. However, the numerical instability of the 3+1 simulation may
pollute the Teukolsky initial data for late extraction times.
Somewhat surprisingly, such a plateau seems to exist even in
situations where a common apparent horizon has yet to form, e.g.,
short-lived evolutions of so-called ``ISCO'' (\emph{Innermost Stable
Circular Orbit}) and ``pre-ISCO'' runs.


Meanwhile, great strides have been made in full 3D simulations over
the last decade, due to the casting of Einstein's equations into more
numerically stable formulations \cite{Nakamura87,Shibata95,
Friedrich96, Frittelli:1996wr, Baumgarte99, Anderson99, Alcubierre00b,
Kidder01a,Sarbach02b,Shinkai02a,Lindblom:2003ad,Bona:2004yp,Pretorius:2004jg},
the development of advanced techniques for handling the singularities
inherent in black-hole space-times \cite{Brandt00, Alcubierre00b,
Alcubierre2003:pre-ISCO-coalescence-times},
and the availability of increased computational resources, coupled
with mesh-refinement techniques
\cite{Imbiriba:2004tp,Fiske:2005fx,Sperhake:2005uf,
Schnetter-etal-03b,Pretorius:2003wc,Pretorius:2005ua}.  The
culmination of these advances is several successful evolutions of
black-hole-binary systems past the symbolic ``one orbit'' barrier
\cite{Bruegmann:2003aw,Pretorius:2005gq}.
For physical systems requiring less than $\sim 100M$ of evolution time
to reach a quiescent final state, these improvements enable the \emph
{direct} extraction of radiation from the 3+1 fields, whether through
Weyl curvature components or Regge-Wheeler-Zerilli variables
\cite{Zerilli70,Abrahams97a}.

These advances, however, do not lessen the relevance of the Lazarus
project. Despite great progress in recent years, it is fair to say
that the problem of black-hole binaries is not completely solved. The
most astrophysically interesting simulations can still not be evolved
long enough to reach their assumed quiescent state. Directly extracted
radiation is still calculated at observer locations that lie in the
``near-field zone'', or may be under-resolved at more distant
locations, and is in general polluted by poor outer-boundary
conditions. As long as such limitations exist, there is a place for
perturbative methods such as Lazarus. Besides, numerical simulations
are still very computationally intensive, and avoiding the last
$\approx 100M$ of binary black-hole evolutions means saving days to
weeks of supercomputer time.




However, Lazarus makes approximations in its approach. Principal among
these is the set of \emph{ad hoc} choices needed to translate the 3+1
curvature information into a Kerr background + perturbations. The
validity of these choices will depend on the data being evolved, and
is difficult to quantify \emph{a priori}.


In this paper, we update the Lazarus project in light of recent work
on transverse frames, in a way that may help resolve some of these
issues. Beetle et al. \cite{Beetle:2004wu} have proposed a way of
identifying the principal directions of a numerical space-time in the
3+1 ADM split. This method --- local in nature --- allows us to narrow
the gap between the numerical tetrad and the Kinnersley tetrad
appropriate to the Teukolsky evolution without any background
assumptions. When calculated with such a tetrad, the longitudinal Weyl
scalars $\Psi_1$ and $\Psi_3$ will vanish, while the ``monopole''
scalar $\Psi_2$ will take on its Kinnersley-tetrad value, and the
transverse scalars $\Psi_0$ and $\Psi_4$, which carry the radiative
degrees of freedom, will differ from their Kinnersley-tetrad values
only by a complex factor. This remaining factor can be compensated
for, on a known Kerr background in Boyer-Lindquist (BL) coordinates,
via a single spin-boost transformation at each point in space.

Although the quasi-Kinnersley frame by no means removes all
uncertainties from the problem of radiative extraction, it goes
sufficiently far that we expect it to improve the Lazarus procedure
considerably. In particular, we expect that the different --- and more
rigorous --- path to Teukolsky waveforms will give us an error
estimate for the tetrad dependence of original results, while the
robustness of the new technique should allow us to attempt consistent
wave extraction from earlier in a numerical evolution.

Additionally, the quasi-Kinnersley frame may achieve much in the
simpler problem of direct radiation extraction
\cite{Beetle:2004wu,Burko:2005fa,Beetle:2002iu,Nerozzi:2004wv,Nerozzi:2005hz}.
In the past \cite{Baker:2002qf}, we used an approximate tetrad to
calculate the Weyl scalar.
Since the quasi-Kinnersley tetrad can be constructed locally, without
knowing Kerr parameters and BL coordinates, we should now achieve a
better approximation to the Kinnersley tetrad during the 3+1
evolution, and thus directly extract waveforms without needing the
background data.


The remainder of this paper is laid out as follows: in Section
\ref{Sec:LazReview}, we summarize the essentials of the original
Lazarus procedure for constructing Cauchy data for Teukolsky
evolution, as well as some of the main results from this procedure. In
Section \ref{Sec:Laz2Theory}, we review the concepts of transverse
frames, and the quasi-Kinnersley frame, and describe the numerical
implementation of these concepts in our evolution code. In Section
\ref{Sec:Results}, we present comparative results from the application
of old and new techniques for three test problems, two of which have
already been addressed with original Lazarus \cite{Baker:2002qf}.
Discussion of the results obtained, and future work can be found in
Section \ref{Sec:Discuss}. Appendix \ref{App_num2Kin} contains
expressions for several quantities related to the evaluation of Weyl
scalars in a numerical tetrad in BL coordinates; Appendix
\ref{App_perturb_res} contains perturbative results for the three test
cases used in Section \ref{Sec:Results}.

\subsection{Notation and Conventions}

In the rest of the paper, we shall generally assume a metric signature
of $( -, +, +, +)$. Our use of algebraic quantities related to the
Kerr-BL solution is non-standard, but consistent with
\cite{Baker:2001sf}; we use an additional quantity $\Lambda=r^2+a^2$ for
compactness.

Our sign convention for the definition of the Weyl scalars is such
that for the Kerr solution in BL coordinates (and using
the Kinnersley tetrad), the only non-vanishing scalar is $\Psi_2 = M /
(r - i \, a \, \cth)^3$.

Vector quantities are denoted by an arrow overhead, except for unit
vectors, which are instead capped by a circumflex
($\hat{\,}$). Complex conjugation is denoted by an overbar
($\bar{\,}$).

\section{The Original Lazarus Method}\label{Sec:LazReview}

Here we provide a summary of the ``original" Lazarus procedure; full
details can be found in \cite{Baker:2001sf}, which we shall refer to
as Paper I from now on.

We start with a full-numerical 3+1 evolution of a space-time of
interest. At some extraction (coordinate) time $T$ --- before the
simulation crashes due to numerical instabilities --- we map the
evolved data to a black-hole perturbative evolution code
\cite{Krivan97a,Pazos-Avalos:2004rp}. This simpler
code can then be evolved stably for as long as is needed to determine
the full history of the gravitational radiation generated.

To assess the level of deviation from Kerr at late times in the 3+1
evolution, \cite{Baker00a} introduced an invariant quantity, the
\emph{speciality index}:
\be
\mathcal{S} = 27 \, \mathcal{J}^2/\mathcal{I}^3,
\ee
where the two complex curvature invariants $\mathcal{I}$ and
 $\mathcal{J}$ are essentially the square and cube of the self-dual
 part, $\mathcal{C}_{a b c d}= C_{a b c d} + (i/2) \, \epsilon_{a b m
 n} \, C^{m n}_{\;\;\;\;\;\;c d}$, of the Weyl tensor:
\be
\mathcal{I}=\mathcal{C}_{a b c d} \, \mathcal{C}^{a b c d} \;\; {\rm and} \;\; 
\mathcal{J}=\mathcal{C}_{a b c d} \, \mathcal{C}^{c d}_{\;\;\;\;m n} \, \mathcal{C}^{m n a b}.
\ee

The geometrical significance of $\mathcal{S}$ is that it measures
deviations from algebraic speciality (in the Petrov classification of
the Weyl tensor). For the unperturbed algebraically special (Petrov
type D) Kerr solution, $\mathcal{S}=1$.  However, for interesting
space-times involving nontrivial dynamics, like distorted black holes,
which are in general not algebraically special (Petrov type I), we
expect $\mathcal{S}=1 + \Delta\mathcal{S}$, and the size of the
deviation $\Delta\mathcal{S} \neq 0$, with leading second perturbative
order, can be used to assess the applicability of black-hole
perturbation theory.

As the expected end-state of most interesting black-hole simulations
is a single spinning (Kerr) hole, the perturbative code implements the
Teukolsky equation \cite{Teukolsky73}. The Kerr metric in BL 
coordinates takes the form:
\bea
ds^2 &=& - \left( 1 - \frac{2 \, M \, r}{\Sigma} \right) \, dt^2 +
\frac{\Sigma}{\Delta} \, dr^2 + \Sigma \, d\theta^2 \nonumber \\
     & & + \frac{\Omega}{\Sigma} \, \ssth \, d\phi^2 - \frac{4 \,
     a \, M \, r}{\Sigma} \, \ssth \, dt \, d\phi,
\eea
where $\Omega \equiv \Lambda \, \Sigma + 2 \, M \, a^2 \, r \, \ssth$,
$\Delta \equiv \Lambda - 2 \, M \, r$, $\Sigma \equiv r^2 + a^2 \,
\ccth$, and $\Lambda \equiv r^2 + a^2$. In these coordinates, the
Teukolsky equation takes the form:
\bea
\label{eqn:Teuk}
&&
\left[{\Lambda^2\over\triangle}-a^2\ssth\right] 
{\partial^2\psi \over \partial t^2}
+{4Mar\over \triangle} {\partial^2 \psi\over\partial t \partial \phi}
 \nonumber\\
&&
+\left[{a^2\over \triangle}-{1 \over\ssth}\right] 
{\partial^2 \psi \over \partial \phi^2}
-\triangle^2 {\partial \over \partial r} \left( {1 \over \triangle} 
{\partial \psi \over \partial r}\right) 
\\
&&
-{1 \over \sth}{\partial \over \partial\theta} 
\left(\sth {\partial \psi \over \partial\theta}\right)
+4 \left[{a\, (r - M)\over \triangle} + i \, \cot\theta\right]
{\partial\psi\over\partial\phi}
 \nonumber\\
&&
+4 \left[{M\, (r^2 - a^2)\over \triangle} - \zeta\right] 
{\partial \psi\over \partial t} 
+(4 \, \cot^2\theta + 2)\psi=0,
 \nonumber
\eea
where $\psi \equiv \rho^{-4} \, \Psi_4$ is the spin-2 \emph{Teukolsky
function}, and $\zeta \equiv r + i \, a \, \cth$. Here $\rho \equiv
m^a \, l_{a;b} \, \bar{m}^b$ is a Newman-Penrose \emph{spin
coefficient}, and $\Psi_4$ is a Newman-Penrose \emph{Weyl scalar},
both calculated using the Kinnersley tetrad.

In the Newman-Penrose formalism \cite{Newman62a}, there are actually
five complex Weyl scalars, formed from contractions of a null tetrad
$(l^a, n^a, m^a, \bar{m}^a)$ with the Weyl tensor:
\bea
\label{eqn:weyl_def}
\Psi_0 & \equiv & C_{a b c d} \, l^a \, m^b \, l^c \, m^d, \nonumber \\
\Psi_1 & \equiv & C_{a b c d} \, l^a \, m^b \, l^c \, n^d, \nonumber \\
\Psi_2 & \equiv & C_{a b c d} \, l^a \, m^b \, \bar{m}^c \, n^d,\\
\Psi_3 & \equiv & C_{a b c d} \, l^a \, n^b \, \bar{m}^c \, n^d, \nonumber \\
\Psi_4 & \equiv & C_{a b c d} \, \bar{m}^a \, n^b \, \bar{m}^c \, n^d. \nonumber
\eea
The $\Psi_i$ encode all the vacuum curvature information of the Weyl
tensor. As space-time scalars, they are coordinate-independent;
however they \emph{do} depend on the particular null tetrad used. With
an appropriate tetrad, in weak-field regions, the interpretation of
the $\Psi_i$ is as follows: $\Psi_2$ embodies the ``monopole"
non-radiative gravitational field; $\Psi_1$ and $\Psi_3$ contain the
longitudinal radiative degrees of freedom (ingoing and outgoing,
respectively), while $\Psi_0$ and $\Psi_4$ contain the physical
transverse radiative degrees of freedom (ingoing and outgoing,
respectively) \cite{Szekeres65}. For a numerical space-time that
contains a Kerr hole plus perturbative gravitational waves, $\Psi_4$
should contain only the appropriate outgoing radiation.

The asymptotic behavior of solutions to the Teukolsky equation is best
expressed in terms of the so-called \emph{tortoise coordinate} $r_*$:
\bea
r_* &=& r + \frac{r_+^2 + a^2}{r_+ - r_-} \, \ln \left| \frac{r -
r_+}{2\,M} \right| - \frac{r_-^2 + a^2}{r_+ - r_-} \, \ln \left|
\frac{r - r_-}{2\,M} \right|, \nonumber\\ r_{\pm} &\equiv& M \pm
\sqrt{M^2 - a^2}.
\eea
The point $r_* = 0$ roughly corresponds to the location of the maximum
of the Kerr solution's scattering potential barrier (see, for example,
Eqn (415) and preceding material in Chap. 8 of
\cite{Chandrasekhar83}). For this reason, it should not be crucial to
obtain initial data for the Teukolsky equation all the way down to the
horizon ($r = r_+ \Rightarrow r_* = - \infty$), as long as we have
data for some $r_* < 0$.

Thus to perform the Teukolsky evolution of radiative data that
corresponds to the late-time evolution of our 3+1 initial data, we
must identify the parameters $(M,a)$ of the Kerr background, and
calculate the radiative Weyl scalar $\Psi_4$ using the Kinnersley
tetrad. Estimation of the physical parameters can be performed fairly
reliably through identification of physical invariants such as the
apparent horizon and ADM mass or correcting (iteratively) the initial
data parameters by the radiative losses. Evaluating $\Psi_4$ using the
correct tetrad is less straightforward.

\subsection{Tetrad Choice}

In BL coordinates, the \emph{Kinnersley null tetrad} takes the form
\cite{Kinnersley_1969}:
\bea
\label{eqn:Kin_tetrad}
\vec{l}_{\rm Kin} &=& \frac{1}{\Delta} \left[ \Lambda, \Delta, 0, a \right], 
\nonumber \\
\vec{n}_{\rm Kin} &=& \frac{1}{2 \Sigma} 
\left[ \Lambda, - \Delta, 0, a \right],\\
\vec{m}_{\rm Kin} &=& \frac{1}{\sqrt{2} \zeta}
\left[ i \, a \, \sth, 0, 1, \frac{i}{\sth} \right]. \nonumber
\eea  
Using this tetrad, the spin coefficient $\rho$ takes the form:
\be
\rho = (r - i \, a \, \cth)^{-1} = 1 / \bar{\zeta}.
\ee

However, the BL coordinates will not, in general, coincide with the
numerical coordinates used in the full 3+1 evolution. We can address
this issue in a post-processing step after the evolution, but we must
still extract enough curvature information to construct $\Psi^{\rm
Kin}_4$. Rather than output all the components of the Weyl tensor
$C_{a b c d}$, it is more efficient to calculate the Weyl scalars with
a numerically convenient tetrad, and transform the results to the
Kinnersley values during post-processing.

The simpler tetrad we use during evolution is a symmetric null tetrad
constructed from the unit hypersurface normal $\hat{\tau}$ and a set
of three orthonormal unit spatial vectors $\hat{e}_{(1)} =
\hat{e}_{\theta}$, $\hat{e}_{(2)} = \hat{e}_{\phi}$, $\hat{e}_{(3)} =
\hat{e}_r$, suitably orthonormalized via a Gram-Schmidt procedure:
\bea
\label{eqn:OT-null_tetrad}
\vec{l}_{\rm num} &\equiv& \frac{1}{\sqrt{2}} ( \hat{\tau} +   \hat{e}_{(3)} )
\;\; , \;\; \vec{n}_{\rm num} \equiv \frac{1}{\sqrt{2}} ( \hat{\tau} -
\hat{e}_{(3)} ),  \nonumber \\
\vec{m}_{\rm num} &\equiv& \frac{1}{\sqrt{2}} ( \hat{e}_{(1)} + i \hat{e}_{(2)}
).
\eea
Similar tetrads have been commonly used in radiation extraction from
3+1 numerical investigations
\cite{Smarr77,Brandt96a,Fiske:2005fx,Sperhake:2005uf}, and such a tetrad was
used in the earliest investigations of the asymptotic radiative degrees of
freedom of the Weyl tensor \cite{Szekeres65}. If we have
long-lived 3D numerical evolutions, whose physical domain extends far
from the strong-field region, the $\Psi_4$ extracted should yield a
good measure of the actual outgoing gravitational radiation. We will
refer to (\ref{eqn:OT-null_tetrad}) hereafter as the \emph{numerical}
tetrad; explicit formulas for the Kerr-BL Kinnersley tetrad are given in
(\ref{eqn:num_tetrad}).

\subsection{Reconstructing Boyer-Lindquist Coordinates}\label{SSec:LazCoordinates}

The reconstruction of the BL coordinates $(t, r, \theta, \phi)$ is
highly non-trivial. We approach the problem via the following \emph{ad
hoc} steps:
\begin{enumerate}
\item assume no polar coordinate distortion;
\item assume that with maximal slicing, numerical time approaches
Boyer-Lindquist time;
\item derive the radial coordinate from the equatorial value of
numerical $\mathcal{I}$;
\item add a radius-dependent correction to the numerical azimuthal
coordinate that zeros out the off-diagonal three-metric component
$\gamma_{r \phi}$.
\end{enumerate}
In short:
\bea
\theta_{\rm BL} &=& \theta_{\rm num}, \label{eqn:num2BL_theta}\\
t_{\rm BL} &=& t_{\rm num}, \label{eqn:num2BL_t}\\
r_{\rm BL} &=& \left. 
\left( 3 \, M^2 / \mathcal{I} \right)^{1/6} 
\right|_{(\theta_{\rm BL} = \pi / 2)}, \label{eqn:num2BL_r}\\
\phi_{\rm BL} - \phi_{\rm num} &=& 
\phi_{\rm offset} = \int_{\infty}^r 
(\gamma_{r' \phi} / \gamma_{\phi \phi}) \, dr'. \label{eqn:num2BL_phi}
\eea

These (last two) coordinate transformations can only be performed as a
post-processing step, after the termination of the full 3D numerical
evolution.

\subsection{Transforming to the Kinnersley Tetrad}

For the Kerr-BL metric, we can move from the numerical tetrad to the
Kinnersley tetrad by a set of linear transformations governed by the
parameters
\be
\label{eqn:params_num2Kin}
A \equiv a \, \sth \sqrt{\frac{\Delta}{\Omega}} \; , \; F_A \equiv
\sqrt{\frac{2 \, \Sigma}{\Delta}} \; , \; F_B \equiv
\frac{\sqrt{\Sigma}}{\zeta}.
\ee
(Note that $|F_B| = 1$, since $\zeta \, \bar{\zeta} = r^2 + a^2 \,
\ccth \equiv \Sigma$.) The specific transformations can be found in
Eqn (\ref{eqn:num2Kin_tetrad}); they consist of a combination of
type-I and type-II tetrad rotations parametrized by $A$, followed by a
type-III ``spin-boost'' transformation parametrized by the real
scaling factor $F_A$ and pure-phase complex factor $F_B$. This
transformation carries over to the Weyl scalars; in particular,
\bea
\label{eqn:num2Kin_Psi4}
\Psi^{\rm Kin}_4 & = & \left[ (D-1)^2 \, 
\Psi^{\rm num}_0 + 4 \, i \, A \, (D-1) \, \Psi^{\rm num}_1 \right. 
\nonumber \\
&& \left. - 6 \, A^2 \, \Psi^{\rm num}_2 - 4 \, i \, A \, (D+1) \,
\Psi^{\rm num}_3 \right. \nonumber \\ && \left. + (D+1)^2 \, \Psi^{\rm
num}_4 \right] / (4 \, F_A^2 \, F_B^2) ,
\eea
where $D \equiv \sqrt{A^2 + 1}$. The equivalent transformations for
the other Kinnersley-tetrad scalars can be found in Eqns
(\ref{eqn:num2Kin_Psi0} - \ref{eqn:num2Kin_Psi3}).

\subsection{Summary of the Original Lazarus Procedure}\label{SSec:Laz_via_num}

To summarize, the original Lazarus procedure involves the following
steps:
\begin{enumerate}
\item construct at every point on the numerical grid a coordinate
null tetrad of the form (\ref{eqn:OT-null_tetrad});
\item calculate the corresponding Weyl scalars $\Psi^{\rm num}_i$, the
tetrad invariants $\mathcal{I}$, $\mathcal{J}$, and the speciality
index $\mathcal{S}$, which we monitor during the entire full-numerical
evolution;
\item determine the BL coordinates from the numerical ones via the
transformations (\ref{eqn:num2BL_theta} - \ref{eqn:num2BL_phi});
\item use the transformation (\ref{eqn:num2Kin_Psi4}) to obtain the
Kinnersley-tetrad $\Psi_4$ and $\partial_t \Psi_4$;
\item evolve the Cauchy data 
$\psi \equiv \rho^{-4} \, \Psi_4$ and $\partial_t \psi$ 
using the Teukolsky equation (\ref{eqn:Teuk});
\item extract the gravitational radiation information, such as 
waveforms and total energy radiated at different extraction 
times $T$. 
\end{enumerate}

\subsection{Results of Original Lazarus}\label{SSec:Laz1Results}

The original Lazarus procedure as outlined above has been applied
extensively to numerical evolutions of black-hole binary data,
extending from head-on collisions, grazing collisions and putative
circular-orbit data at various orbital separations
\cite{Baker00b,Baker:2001nu,Baker:2002qf}.



One of the main shortcomings of Lazarus is the \emph{ad hoc} nature of
the coordinate transformations in (\ref{eqn:num2BL_theta} -
\ref{eqn:num2BL_phi}). Even with data in the linearized regime,
Lazarus will be sensitive to how well the numerical space-time
satisfies these coordinate assumptions. For instance, we have no
guarantee that there is no angular distortion between the numerical
and BL radial coordinates (although studies of post-merger apparent
horizons have noted a definite tendency of the horizon shape to
circularize in numerical coordinates).

An even bolder assumption is that maximal slicing will yield a
late-time lapse with the Kerr-BL shape. The Kerr-BL lapse satisfies
the maximal slicing equation; however, the numerical lapse also
depends on the boundary conditions. In practice, we use Dirichlet
boundary conditions, with values equal to that of the Schwarzschild
lapse ($M = 1$) at the same coordinate position. This choice will give
us a lapse shape (and hence numerical time) \emph{qualitatively} like
Kerr-BL; however we only have experimental \emph{quantitative}
experience with the quality of the fit. It can be shown for instance
that the numerically obtained lapse quickly takes the form of the BL
lapse over points in the exterior of the horizon; the greatest
deviations occur within the ``potential barrier'' at $r_*$. A plot
demonstrating this for evolved QC0 binary data is shown in Fig. 6 of
Paper I.

A complete and unambiguous solution of the coordinate problem is not
yet available. Until it is, it seems sensible to work to minimize
Lazarus's coordinate dependence. One obvious area to address is the
transformation (\ref{eqn:num2Kin_Psi4}) of the curvature from
numerical to Kinnersley tetrad; the numerical Weyl scalars are both
mixed and scaled by coordinate-dependent factors. An improvement has
been made possible by evaluating the Weyl scalars instead with the
\emph{quasi-Kinnersley tetrad}, which is the subject of the next
section.

\section{Lazarus with the Quasi-Kinnersley Frame}\label{Sec:Laz2Theory}

In this section, we review some recent results on transverse frames in
numerical space-times, and discuss how they can be used to develop new
techniques to improve the Lazarus procedure described in Section
\ref{Sec:LazReview}. The main aim is to move closer to the calculation
of the actual Kinnersley tetrad during the full 3+1 evolution, and
thus to minimize the need for \emph{ad hoc} coordinate and tetrad
correction schemes.

\subsection{Transverse Frames and the Quasi-Kinnersley Frame}
\label{SSec:TransFrames}

The Kinnersley tetrad on a Kerr background is \emph{transverse} --
when calculated with this tetrad, the ``longitudinal'' Weyl scalars
$\Psi_1$ and $\Psi_3$ vanish. This continues to hold for the
non-trivial situation of a Kerr hole plus perturbing radiation.
However, the coordinate and slicing ambiguity of a generic numerical
space-time, even when only perturbatively different from Kerr, mean
that this tetrad can be difficult to identify. The Lazarus
tetrad-transformation procedure outlined in the previous section will
\emph{not}, in general, yield a transverse tetrad.

A way of identifying transverse tetrads is through finding the
eigen-bivectors of the self-dual Weyl tensor $\mathcal{C}_{a b c d}$
\cite{Beetle:2004wu} (see also Chapter 4 of \cite{Kramer80}). When
expressed in a 3+1 decomposition, the Weyl tensor is projected to
$C_{a c} \equiv \mathcal{C}_{a b c d} \, \tau^b \, \tau^d$, and the
eigen-equation becomes
\be
\label{eqn:4d_e-eqn}
C^a_{\;\;c} \, \sigma^c \equiv (E^a_{\;\;c} - i \, B^a_{\;\;c}) \,
\sigma^c = \lambda \, \sigma^a,
\ee
where $E^a_{\;\;b}$ and $B^a_{\;\;b}$ are the so-called
\emph{electric} and \emph{magnetic} parts of the Weyl tensor
\cite{Matte53,Smarr77}. This can be recast by projecting the Weyl
tensor onto the \emph{orthonormal triad} $\{\hat{e}_{(i)}\}$:
following \cite{Kramer80}, we can write (summation is implied over
parenthetical indices):
\bea
C_{a c} &=& Q_{(i)(j)} \, e_{(i) a} \, e_{(j) c},\\
\sigma^a &=& V_{(i)} \, e_{(i)}^a, \label{eqn:sig_def}
\eea
where $Q_{(i)(j)}$ is a symmetric complex $3 \times 3$ matrix whose
components are:
\bea
\label{eqn:reduced_weyl}
Q_{(1)(1)} &=& - \Psi_2 + ( \Psi_0 + \Psi_4 ) / 2, \nonumber\\
Q_{(1)(2)} &=& i ( \Psi_0 - \Psi_4 ) / 2 \;\; , \;\; Q_{(1)(3)} =
\Psi_3 - \Psi_1, \nonumber\\ Q_{(2)(2)} &=& - \Psi_2 - ( \Psi_0 +
\Psi_4 ) / 2,\\ Q_{(2)(3)} &=& - i ( \Psi_1 + \Psi_3 ) \;\;,\;\;
Q_{(3)(3)} = 2 \Psi_2, \nonumber
\eea
and the $\Psi_i$ are calculated using the symmetric tetrad
(\ref{eqn:OT-null_tetrad}). Then the eigen-equation
(\ref{eqn:4d_e-eqn}) reduces to
\be
\label{eqn:3d_e-eqn}
Q_{(i)(j)} \, V_{(j)} = \lambda \, V_{(i)}.
\ee

The three eigenvalues $\lambda$ of the matrix $Q_{(i)(j)}$ label three
transverse frames; the corresponding eigenvectors give the principal
directions of the space-time. One of the $\lambda$ is preferred --- it
will be analytic near the point $\mathcal{S} = 1$ in the complex
$\mathcal{S}$-plane. The frame related to this $\lambda$ is called the
\emph{quasi-Kinnersley frame}. The preferred $\lambda$ will be
numerically equal to $2 \times \Psi^{\rm Kin}_2$, the value of the
monopole Weyl scalar as calculated with the Kinnersley tetrad.

Beetle et al. \cite{Beetle:2004wu} have described the analytic
determination of the Weyl eigenvalues. In practice, establishing
analyticity of the the eigenvalues is not necessary numerically --
Mars \cite{Mars_2001} has pointed out that close to Kerr, the
eigenvalue with the largest complex norm will give the desired frame;
this conclusion has been made more secure by \cite{Beetle:2004wu}, who
have shown that this is a valid conclusion everywhere in the disc $\|
\mathcal{S} - 1\| < 1$. Instead of following the analytic route,
therefore, we use the LAPACK routine {\tt zgeev}~\cite{lapack} to
determine numerically the eigenvalues and eigenvectors of
$Q_{(i)(j)}$. We select the largest-modulus eigenvalue as the
appropriate (quasi-Kinnersley) one.

\subsection{Tetrad Reconstruction}\label{SSec:NewTetrad}

Beetle et al. \cite{Beetle:2004wu} lay out a procedure for
constructing a tetrad $(l^a, n^a, m^a, \bar{m}^a)_{\rm qK}$ in the
quasi-Kinnersley frame from the eigenvector $V_{(i)}$: first construct
$\sigma^a$ from (\ref{eqn:sig_def}); normalize it so that
$\vecnorm{\sigma} = 1$; separate into real and imaginary parts:
$\sigma^a = x^a + i \, y^a$; construct a third orthogonal vector $z^a
= \varepsilon^{a}_{\;\; b c} x^b y^c$; then the new null tetrad
vectors are:
\bea
\vec{l}_{\rm qK} &=& \frac{|c|}{\sqrt{2}} \left( \vecnorm{x} \, \hat{\tau} + \frac{\vec{x} + \vec{z}}{\vecnorm{x}} \right),\label{eqn:lT_construct}\\
\vec{n}_{\rm qK} &=& \frac{|c|^{-1}}{\sqrt{2}} \left( \vecnorm{x} \, \hat{\tau} + \frac{- \vec{x} + \vec{z}}{\vecnorm{x}} \right),\label{eqn:nT_construct}\\
\vec{m}_{\rm qK} &=& \frac{e^{i \, \chi}}{\sqrt{2}} \left( \sqrt{\vecnorm{x}^2 - 1} \, \hat{\tau} + \frac{(\vec{z} - i \vec{y})}{ \sqrt{\vecnorm{x}^2 - 1} } \right) \label{eqn:mT_construct},
\eea
where $c \equiv |c| \, e^{i \, \chi}$ is an arbitrary spin-boost
parameter. We take $|c| = 1, \chi = \pi/2$, in order to produce a
tetrad that asymptotes to the original numerical tetrad at large
distances. The subscript ``qK'' will refer to this specific choice of
$c$ from this point.

If it happens that the eigenvector $\sigma^a$ is identically real,
$\vecnorm{x} = 1$, $\vecnorm{y} \sim \vecnorm{z} \sim
\sqrt{\vecnorm{x}^2 - 1} = 0$ (see Appendix Section
\ref{App_QC0_perturb} for analytic examples in the ``close-slow''
limit of binary Bowen-York data). This is not a problem when
constructing $\vec{l}_{\rm qK}$ and $\vec{n}_{\rm qK}$; however Eqn
(\ref{eqn:mT_construct}) will now be undefined. A valid complex null
vector $\vec{m}_{\rm qK}$ can still be formed in this case, by
replacing the second term in parentheses by any linear combination
$\vec{a} + i\,\vec{b}$ of two unit spatial vectors orthogonal to
$\vec{x}$. Beetle et al. supply one such choice in Eqn (29) of
\cite{Beetle:2004wu}. However, the resulting $\vec{m}_{\rm qK}$ at
this point will have real and imaginary components that may not match
continuously to neighboring points.

In a continuous domain, we could imagine evaluating $\vec{m}_{\rm qK}$
by taking a limit from neighboring points; on a numerical domain
(especially in 3D), this is an impractical approach, as (i) it would
necessitate knowing in advance which points would need to be
interpolated, and (ii) we would need a very dense numerical mesh to
carry out such an interpolation. Additionally, there are cases where
the pathological points are not isolated, but cover the entire
domain. This is the case for Schwarzschild and Brill-Lindquist
data. In such cases, no interpolation procedure is possible.

For these reasons, we use an alternative tetrad reconstruction
procedure, one that avoids pathologies entirely. We start by following
the prescription of Eqns (\ref{eqn:lT_construct} -
\ref{eqn:nT_construct}) for the reconstruction of the real null
vectors $\vec{l}_{\rm qK}$ and $\vec{n}_{\rm qK}$, as in
\cite{Beetle:2004wu}; these vary smoothly from point to point, even
for pathological regions when $\vecnorm{z} = 0$.

Next, we take the original numerical complex null vector and split it
into real and imaginary parts: $\vec{m} \equiv \vec{X} + i
\vec{Y}$. Now starting from these vectors, orthonormalize them
according to a set of Gram-Schmidt-like steps. Since $\vec{l}_{\rm
qK}$ and $\vec{n}_{\rm qK}$ are already correctly orthonormalized, the
remaining requirements are:
\bean
\vec{m}_{\rm qK} \cdot \vec{l}_{\rm qK} &=& 0 \Rightarrow \vec{X} \cdot \vec{l}_{\rm qK} = 0,\;\;  \vec{Y} \cdot \vec{l}_{\rm qK} = 0,\\
\vec{m}_{\rm qK} \cdot \vec{n}_{\rm qK} &=& 0 \Rightarrow \vec{X} \cdot \vec{n}_{\rm qK} = 0,\;\;  \vec{Y} \cdot \vec{n}_{\rm qK} = 0,\\
\vec{m}_{\rm qK} \cdot \vec{m}_{\rm qK} &=& 0 \Rightarrow \vec{X} \cdot \vec{Y} = 0,\;\; \vec{X} \cdot \vec{X} = \vec{Y} \cdot \vec{Y},\\
\vec{m}_{\rm qK} \cdot \vec{\bar{m}}_{\rm qK} &=& 1 \Rightarrow \vec{X} \cdot \vec{X} + \vec{Y} \cdot \vec{Y} = 1.
\eean
The last two equations combine to imply that:
\[
\vec{X} \cdot \vec{X} = \vec{Y} \cdot \vec{Y} = 1/2.
\]

To impose these conditions, we begin with $\vec{X} = \hat{e}_{(1)} /
\sqrt{2}$, and enforce the conditions in turn (note that since
$\vec{l}_{\rm qK}$ and $\vec{n}_{\rm qK}$ are null, the Gram-Schmidt
procedure looks slightly unusual):
\bean
\vec{X} &\rightarrow& \vec{X} + (\vec{X} \cdot \vec{l}_{\rm qK}) \, \vec{n}_{\rm qK},\\
\vec{X} &\rightarrow& \vec{X} + (\vec{X} \cdot \vec{n}_{\rm qK}) \, \vec{l}_{\rm qK},\\
\vec{X} &\rightarrow& \vec{X} / \sqrt{2 (\vec{X} \cdot \vec{X})}.
\eean
In a similar manner, we take $\vec{Y} = \hat{e}_{(2)} / \sqrt{2}$, and
enforce orthogonality to $\vec{X}$ just before normalization.

It can be seen that the real and imaginary parts of
(\ref{eqn:mT_construct}), used as $\vec{X}$ and $\vec{Y}$, pass
untouched through the Gram-Schmidt steps above. The choice we make of
beginning with the \emph{original} $\vec{m}$ instead will mean that
our final $\vec{m}_{\rm qK}$ differs from (\ref{eqn:mT_construct}) by
no more than a spin term. Unlike (\ref{eqn:mT_construct}), the
Gram-Schmidt procedure described can be used everywhere in the
numerical domain, and guarantees a smooth behavior moving between
neighboring non-pathological and pathological points.

Following the procedure as outlined here exactly reproduces the
quasi-Kinnersley tetrad for Kerr-BL, and behaves well for
Brill-Lindquist and Bowen-York-type binary data.

\subsection{Spin-Boost Fixing and the Kinnersley Tetrad}\label{SSec:SpinBoost}

The tetrad obtained from this procedure will be transverse, and
moreover, will be in the same transverse frame as the Kinnersley
tetrad --- differing only by a Type-III --- or \emph{spin-boost} --
transformation. Such a transformation will leave the scalar $\Psi_2$
unchanged, but will have a strong effect on the radiative fields
$\Psi_0$ and $\Psi_4$, scaling and mixing polarizations.

Lacking an unambiguous and natural way to lock down the spin-boost
needed to obtain the Kinnersley tetrad from the quasi-Kinnersley frame
member, we return to our knowledge of the Kerr background. It can be
shown that for Kerr-BL, the null tetrad produced from the correct
transverse eigenvector (by following the construction in
\cite{Beetle:2004wu}) is
\bea
\vec{l}_{\rm qK} &=& \frac{1}{\sqrt{2 \Delta \Sigma}}\left[ \Lambda, \Delta, 0, a \right], \nonumber \\
\vec{n}_{\rm qK} &=& \frac{1}{\sqrt{2 \Delta \Sigma}}\left[ \Lambda, - \Delta, 0, a \right],\\
\vec{m}_{\rm qK} &=& \frac{- i}{\sqrt{2 \Sigma}} \left[ i a \sth, 0, 1 , \frac{i}{\sth} \right] \nonumber,
\eea
that is, the new tetrad is related to the Kinnersley tetrad
(\ref{eqn:Kin_tetrad}) via:
\bea
\vec{l}_{\rm Kin} &=& \sqrt{\frac{2 \Sigma}{\Delta}} \vec{l}_{\rm qK} \;\;,\;\; \vec{n}_{\rm Kin} = \sqrt{\frac{\Delta}{2 \Sigma}} \vec{n}_{\rm qK}, \nonumber\\
\vec{m}_{\rm Kin} &=& \frac{\bar{\zeta}}{\sqrt{\Sigma}} \vec{m}_{\rm qK}.
\eea
This implies that the corresponding Kinnersley-tetrad $\Psi_4$ can be
obtained simply from:
\be
\label{eqn:new_weyl_transform}
\Psi^{\rm Kin}_4 = \Psi^{\rm qK}_4 / F_A^2 \, F_B^2.
\ee
Note that this transformation, unlike (\ref{eqn:num2Kin_Psi4}) for original
Lazarus, does not involve any mixing of the Weyl scalars.

\subsection{Summary of the New Procedure}\label{SSec:Laz_via_QK}

Thus the new Lazarus procedure involves the following sequence of
steps: 
\begin{enumerate}
\item construct at every point on the numerical grid a coordinate
null tetrad of the form (\ref{eqn:OT-null_tetrad});\label{it:step1}
\item calculate the corresponding Weyl scalars $\Psi^{\rm num}_i$,
the tetrad invariants $\mathcal{I}$, $\mathcal{J}$, and the speciality
index $\mathcal{S}$, which we monitor during the full-numerical
evolution;
\item construct the matrix $Q_{(i)(j)}$ (\ref{eqn:reduced_weyl}), and
find its eigenvalues and eigenvectors;
\item pick the eigenvalue $\lambda$ with the largest (complex)
magnitude;
\item use the corresponding unit-norm eigenvector $\vec{V}$ to form
the principal (complex) spatial vector $\sigma^a \equiv V_{(i)} \,
e^a_{(i)}$;
\item construct a new null tetrad according (\ref{eqn:lT_construct}),
(\ref{eqn:nT_construct}), and the pathology-avoiding steps following;
\item recalculate the Weyl scalars with this new tetrad;\label{it:step7}
\item determine the BL coordinates from the numerical ones via the
transformations (\ref{eqn:num2BL_theta} - \ref{eqn:num2BL_phi});
\item use the transformation (\ref{eqn:new_weyl_transform}) to produce
the Kinnersley-tetrad $\Psi_4$ and $\partial_t
\Psi_4$;\label{it:step8}
\item evolve the Cauchy data $\psi \equiv \rho^{-4} \, \Psi_4$ and
$\partial_t \psi$ using the Teukolsky equation (\ref{eqn:Teuk});
\item extract the gravitational radiation information, such as 
waveforms and total energy radiated at different extraction 
times $T$.
\end{enumerate}

In the above, only steps \ref{it:step1}-\ref{it:step7} can be carried
out in the full-numerical evolution; step \ref{it:step8} is a
post-processing step, as it still involves knowledge of the
appropriate Kerr background parameters ($M$,$a$), and BL
coordinates. While the new Lazarus approach outlined above does not
solve this problem, it reduces the dependence of the Teukolsky data on
the tetrad choices, by reducing the number of functions that
depend on the BL coordinates.


\section{Numerical Evolutions}\label{Sec:Results}

We have applied the new transverse-frame-based techniques outlined
above to three numerical regimes of increasing complexity: Bowen-York
data for one spinning hole, Brill-Lindquist data for the head-on
collision of two black holes, initially at rest, and transversely
boosted binary Bowen-York data, representing an ``Innermost Stable
Circular Orbit''. Two of these data sets were already investigated in
detail using the existing Lazarus approach in \cite{Baker:2002qf}; our
new techniques should allow some additional estimate on their
validity.

The results have been obtained from 3D evolutions performed with the
standard ADM decomposition of Einstein's equations \cite{York79},
together with zero shift and maximal slicing lapse. It may be objected
that this combination of evolution systems and gauge conditions is not
state-of-the-art. More sophisticated and stable systems are in wide
use at present, and we would certainly expect them to produce much
longer-lived stable full 3D evolutions. In particular, a new
numerical-relativity framework ({\it LazEv}) has been successfully
developed \cite{Zlochower:2005bj} and is currently available to the
authors of this paper; it allows for evolutions of black-hole binaries
using higher-order finite differencing, with the BSSN formulation of
Einstein's equations and dynamical gauge conditions. Nevertheless, we
have two reasons for using the ``ADM + maximal slicing + zero shift''
combination: First, this combination was that used for the original
Lazarus tests; if we desire a fair comparison of old and new methods,
focused on the new tetrad methods we have developed, use of more
modern and stable evolution methods will only obscure the
results. Second, the introduction of more sophisticated gauges will
necessitate the reconsideration of the the numerical-to-Kerr-BL
coordinate transformation. In particular, the use of other lapse
evolution schemes may lead to a lapse shape significantly different to
the maximal and Kerr-BL shapes. Also, the presence of a non-zero shift
vector will alter the azimuthal-angle relationship
(\ref{eqn:num2BL_phi}).

We carried out the coding and testing of the concepts presented here
using the Cactus \cite{Cactusweb} framework. Post-processing of the 3D
data and subsequent Teukolsky evolution was done with stand-alone
codes.

In these evolutions, we were also monitoring the appearance of a
single merged black hole using the apparent horizon finder {\tt
AHFinderDirect} \cite{Thornburg2003:AH-finding}.

%

\subsection{Spinning Bowen-York Data}

The Bowen-York solution \cite{Bowen80} representing a single black
hole with angular momentum $J$ should resemble --- once the
Hamiltonian constraint has been solved --- a single Kerr black hole
with gravitational radiation on top. The rotational symmetry of the
solution about its spin axis (the coordinate polar or $z$ axis) means
that only radiation with $m = 0$ will be present, and that there
should be no net loss of angular momentum through gravitational
radiation. Some analytic treatment of this data is presented in the
Appendix Section \ref{App_BYSpin_perturb}, based on Gleiser et al.'s
perturbative analysis \cite{Gleiser:1998ng}.

In an early numerical investigation of the suitability of such data
for Lazarus, we evaluated this data --- using a full-numerical
solution to the constraints --- and extracted the radiation content of
the initial data via the new Lazarus method. We found that for small
spins ($J / M^2 \lesssim 0.2$), a spatial resolution of up to $M / 80$
would be necessary to identify unambiguously the leading-order real
($\ell = 2$) \emph{and} imaginary ($\ell = 3$) parts of the Teukolsky
function. Such a resolution is not feasible for a unigrid run, if the
outer boundaries are to be outside the strong-field region. For this
reason, we have used a larger initial spin for full-numerical
simulations.

This data was implemented on the initial numerical slice in Cactus,
with bare mass $m = 0.858 \, M_{\rm ADM}$, and spin angular momentum
$J = 0.553 \, M_{\rm ADM}^2$; after numerical solution of the
Hamiltonian constraint, the ADM mass of the initial data was $M_{\rm
ADM} = 1.0$. To extend the physical domain, we used a ``transition
fisheye'' radial transformation of the numerical
coordinates~\cite{Baker:2001sf}. The explicit form we used in this
paper can be found in Eqn (114) of \cite{Alcubierre02a}, with
parameters $(a = 7,s = 2,r_0 = 7)$; this produced a physical radial
extent of $33.6 \, M$ from a numerical radial extent of $10.79 M$. We
used three spatial resolutions --- $M / 12$, $M / 18$ and $M /
24$. Because the solution \emph{below} the $x$-$y$ plane is trivially
related to that \emph{above} the plane, we were able to evolve the
upper bitant only.  The 3+1 simulation died due to numerical
instabilities before $25 \, M$ of coordinate time at the coarsest
resolution, before $19 \, M$ at the medium resolution, and before $18
\, M$ at the finest resolution.

\begin{figure}
  \begin{center}
    \includegraphics*[width=3.3in]{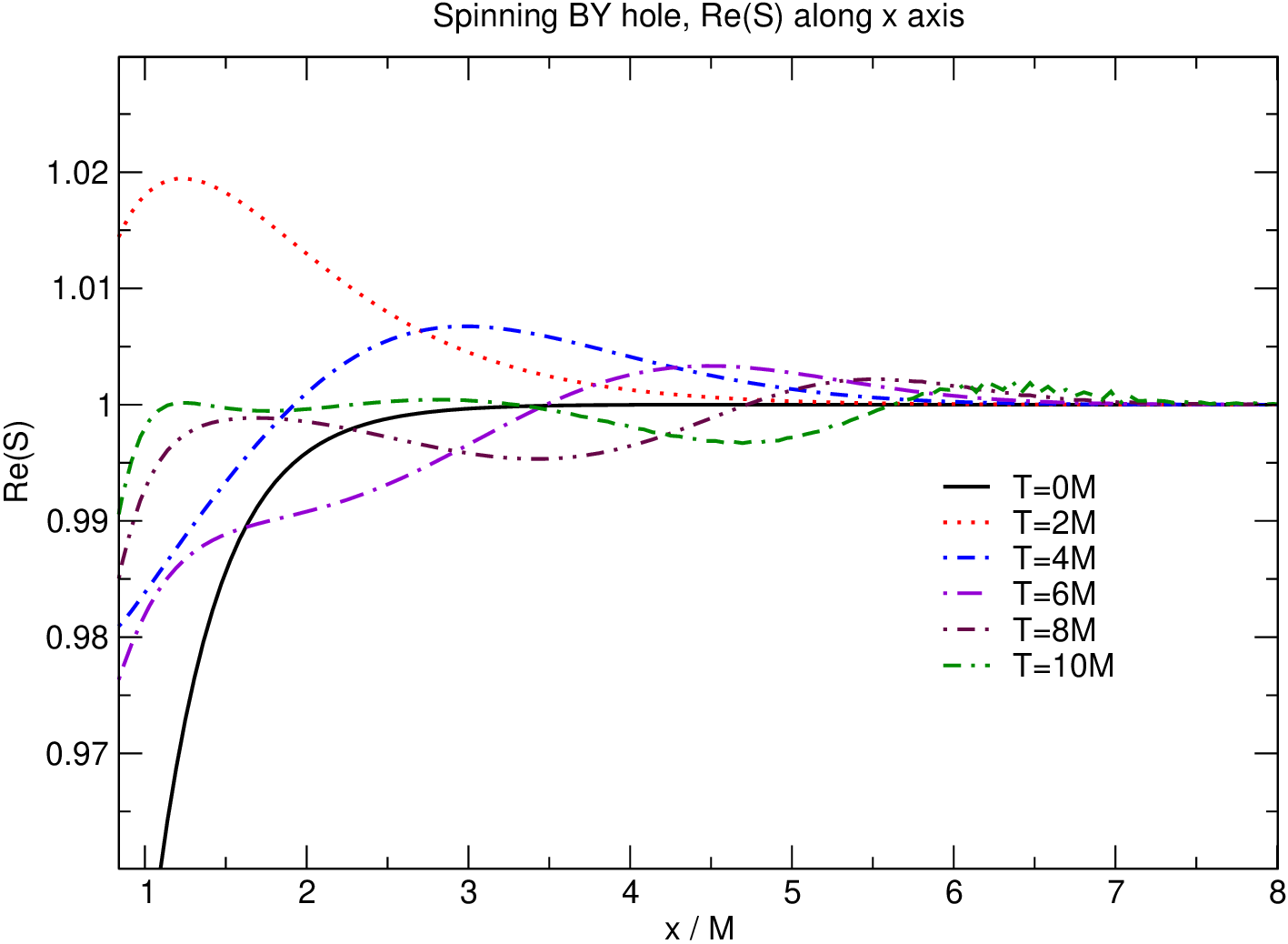}
  \end{center}
  \caption{The real part of the $\mathcal{S}$ invariant along the $x$
    axis for Spinning BY data, at different extraction times
    $T$.}
  \label{fig:BY-Sx}
\end{figure}

For such a weak-field case, we expect the Lazarus techniques to be
applicable from a very early time. An apparent horizon is already
present in the initial data, and Fig. \ref{fig:BY-Sx} shows the value
of the (real part of the) speciality index $\mathcal{S}$ along the $x$
axis outside the horizon, evaluated at several times in the
evolution. At all times after the initial slice, the deviation from
the Kerr value of $\mathcal{S} = 1$ is at most $\sim 2 \%$. Thus the
spinning BY data should certainly be within the linearized regime
almost immediately. (Nevertheless, Lazarus results will be sensitive
to how well the numerical space-time satisfies the coordinate
assumptions of Section \ref{SSec:LazCoordinates}.)

\begin{figure}
  \begin{center}
    \includegraphics*[width=3.3in]{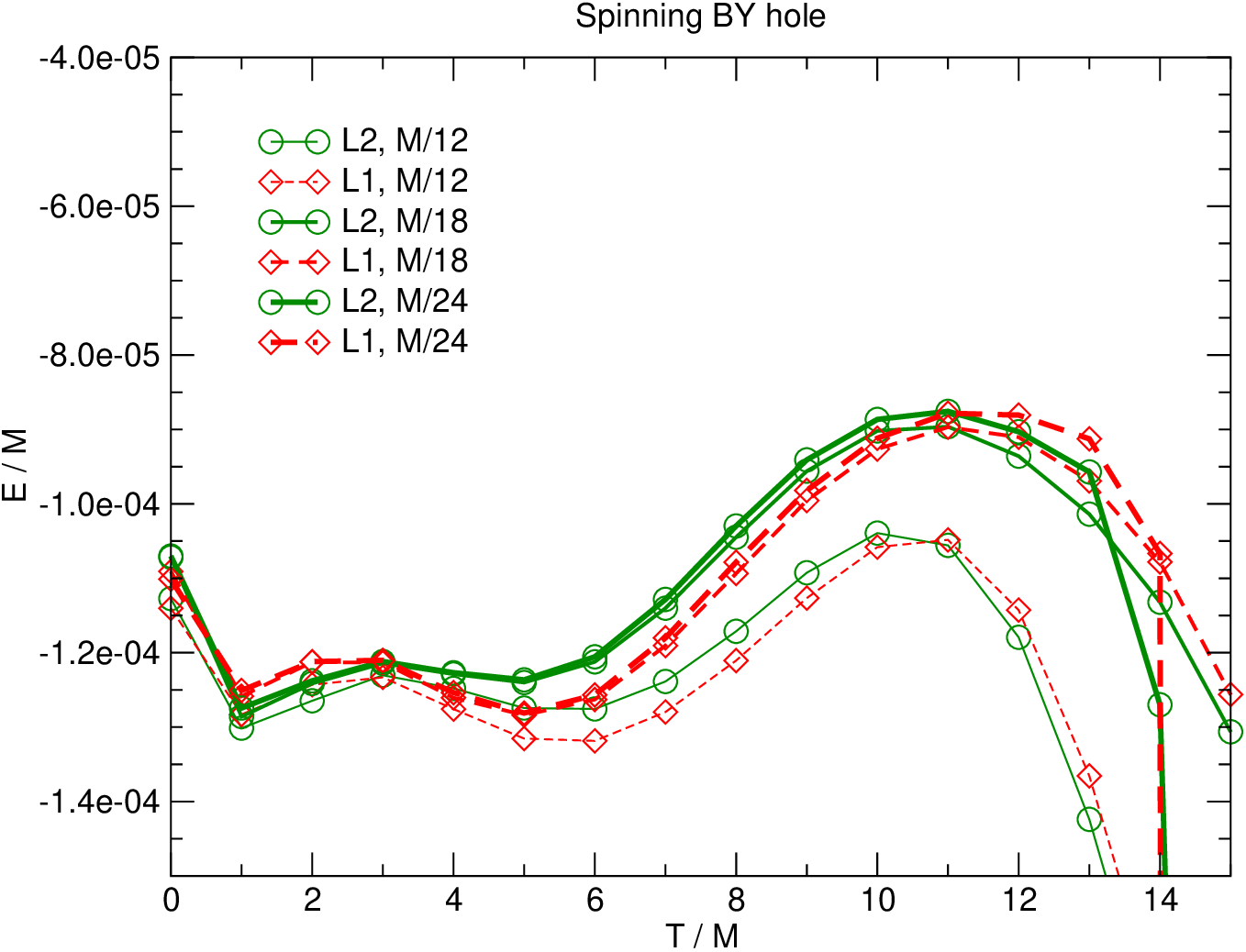}
  \end{center}
  \caption{Total radiated energy for Spinning BY data as calculated
  from original (L1) and new (L2) Lazarus methods, as a function of
  extraction time $T$.}
  \label{fig:BY-EvsT}
\end{figure}

We analyzed the evolved data with both the original and new Lazarus
approaches. At every $M$ of coordinate time evolution, the Weyl
scalars were calculated using both the original and new tetrads, then
mode-decomposed and saved to file. We post-processed this data to
produce the Teukolsky Cauchy data $(\psi, \partial_t \psi)$, which was
then evolved using a code that solves the Teukolsky Equation. The
total extracted energy as a function of extraction time $T$ is shown
in Fig. \ref{fig:BY-EvsT}. Between extraction times of $1M$ and $5M$,
we see an emitted-energy plateau at around $1.25 \times 10^{-4} \, M$,
common to both original and new Lazarus, with a deviation of $\sim 3.5
\%$ between old and new values.

We would expect that once we have entered an energy plateau, we should
remain there, at least until the radiation has left the outer boundary
of the full-numerical domain. However, for later extraction times,
from $6M$ to $10M$, we see a monotonic \emph{decrease} in emitted
energy, leading to a low point of $\sim 0.88 \times 10^{-4} \, M$, a
drop of $\sim 30 \, \%$. This degradation is common to both original and
new Lazarus.


This behavior may be an artifact of the underlying ADM evolution
system, or other aspects of the Lazarus procedure; even for the
relatively high spin chosen here, spinning Bowen-York is a weak source
of gravitational radiation, and presumably extremely sensitive to
details in the simulation. Further investigation with newer, more stable
and accurate, evolution systems may illuminate the situation.

The level of emitted energy is consistent with the full-numerical
results shown in ~\cite{Dain:2002ee} and the perturbative results of
Gleiser et al. \cite{Gleiser:1998ng}: they obtain a total emitted
energy of $\sim 2.5 \times 10^{-4} \, M$ for a spin of $J_{\rm ADM} =
0.55 \, M^2$. Although significantly larger than our peak energy of
$\sim 1.25 \times 10^{-4} \, M$, it falls into the ``factor of two''
accuracy the authors assess for their method at high spins.

\begin{figure}
  \begin{center}
    \includegraphics*[width=3.3in]{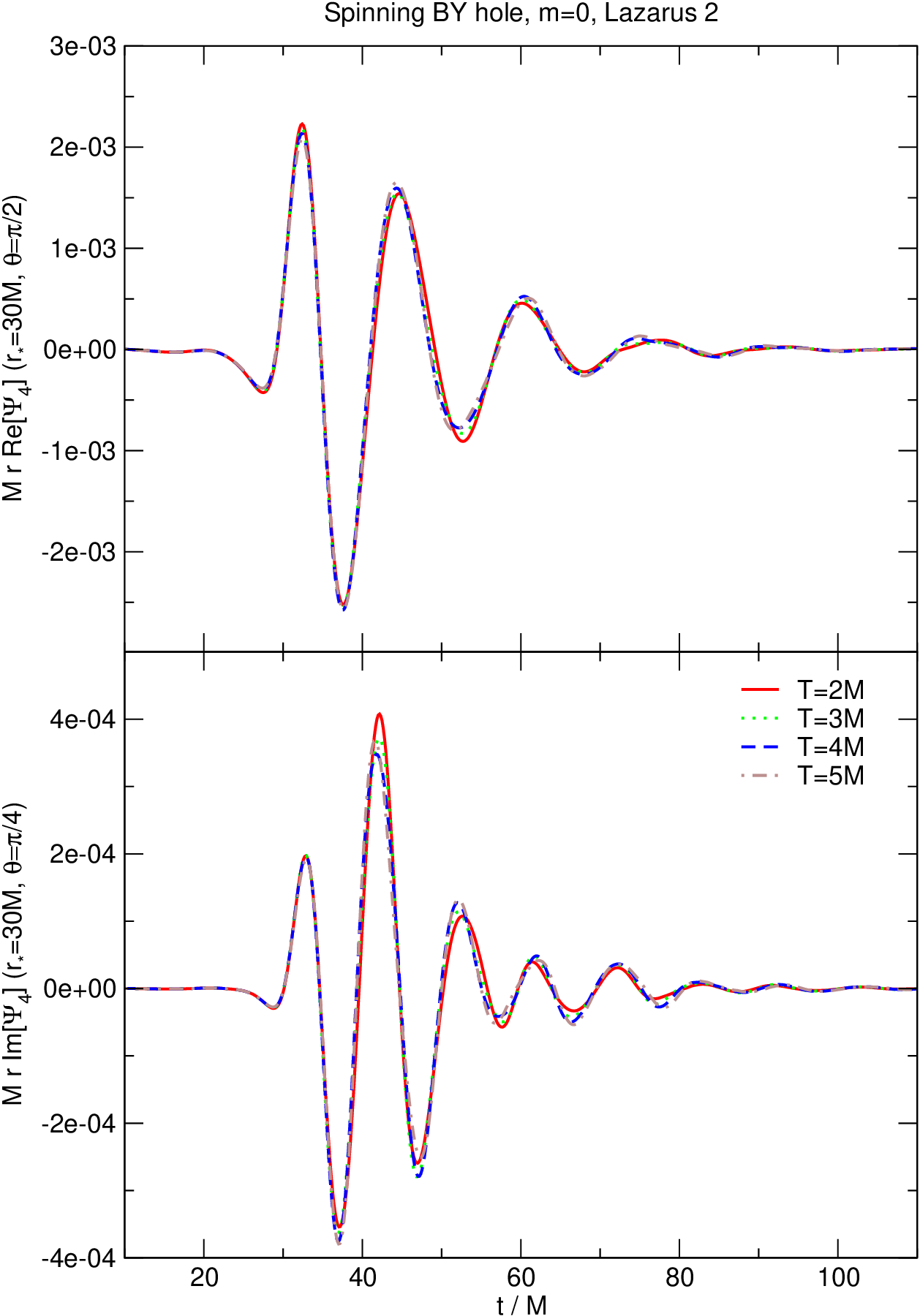}
  \end{center}
  \caption{Evolution of the $m=0$ Spinning BY waveform for the new
     (L2) Lazarus method, evaluated at $r_* = 30\,M$ along maximal
     amplitude directions: $\theta = 90^\circ$ for $Re(\Psi_4)$,
     $\theta = 45^\circ$ for $Im(\Psi_4)$. The five curves represent
     five different extraction times in the energy plateau of
     Fig. \ref{fig:BY-EvsT}.}
  \label{fig:BY-WF-L2}
\end{figure}

Selected Teukolsky-evolved waveforms (the rescaled Weyl scalar
$\Psi_4$), measured at $(r_* = 30\,M, \theta = 0)$, are shown in
Fig. \ref{fig:BY-WF-L2}. The waveforms shown are for the new tetrad
method only (old-method waveforms are indistinguishable). There is
good agreement especially between extraction times $T = 3\,M$ and
$5\,M$. For later extraction times, the waveforms lose coherence
after a few wavelengths.

%

\subsection{Brill-Lindquist Data}

This test revisits the case of head-on collisions of Brill-Lindquist
data, treated previously in the Lazarus context
\cite{Baker00b,Baker:2002qf}. Data of this type was studied in a
perturbative form in Ref.~\cite{Abrahams95c}; 
based on the full-numerical results available at the time,
perturbative results overestimate the total emitted energy for large
separations. Qualitatively similar conclusions can be drawn using
Misner initial data~\cite{Lousto99a}.

The specific initial data chosen has two holes with equal bare masses
$m = 0.50\,M$, with a coordinate separation along the $y$ axis of
$2.303 \, M$, yielding a proper horizon-to-horizon separation of $L =
4.9M$, and no linear or angular momentum (this data set was referred
to as ``$P = 0$'' in \cite{Baker:2002qf}). The symmetry of the problem
allows us to use only the first octant of the numerical grid during
evolution. Otherwise, the grid extent and fisheye transformation were
identical to those for the spinning Bowen-York case above.

\begin{figure}
  \begin{center}
    \includegraphics*[width=3.3in]{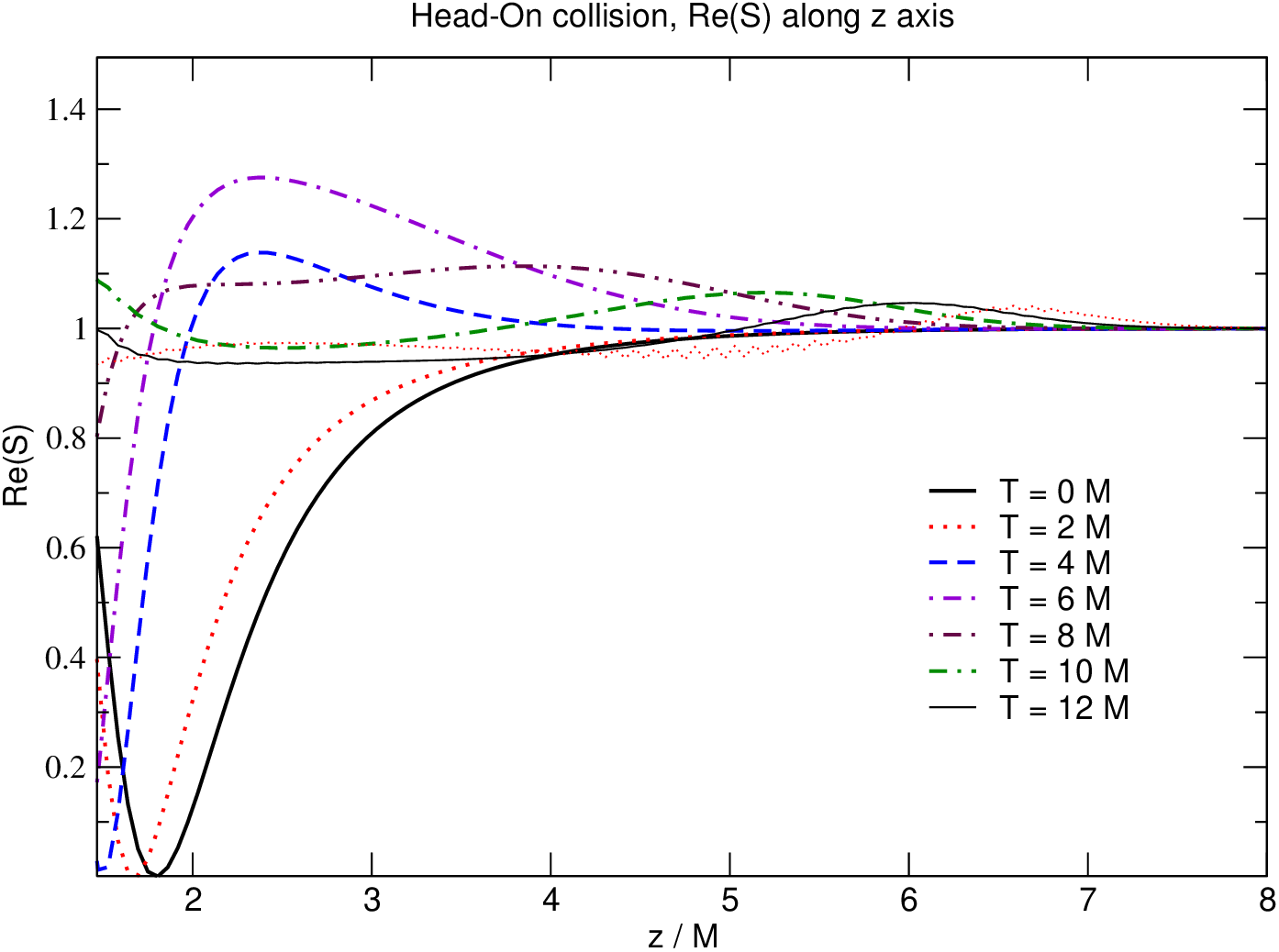}
  \end{center}
  \caption{The $\mathcal{S}$ invariant along the $z$ axis for Head-On
  data, at different extraction times $T$.}
  \label{fig:HO-Sz}
\end{figure}

Again, we evolved the initial data using the ``standard ADM''
evolution scheme with maximal slicing and zero shift. The finest
resolution run crashed due to numerical instabilities before $20 M$ of
coordinate time. Nevertheless, a common apparent horizon was found at
coordinate time $t \approx 8 \, M$, and there is reason to believe
that a common event horizon is present from $t \approx 3 \, M$
\cite{Diener_comm} and a continuous potential barrier surrounding the
strong-field region around even earlier. We show the evolution of the
speciality index $\mathcal{S}$ in Fig. \ref{fig:HO-Sz}. While
deviations from unity are always large near the punctures, the overall
deviation outside the eventual common horizon location have dropped to
below $10 \%$ by $T \approx 8 \, M$, indicating we have entered the
linear regime.

\begin{figure}
  \begin{center}
    \includegraphics*[width=3.3in]{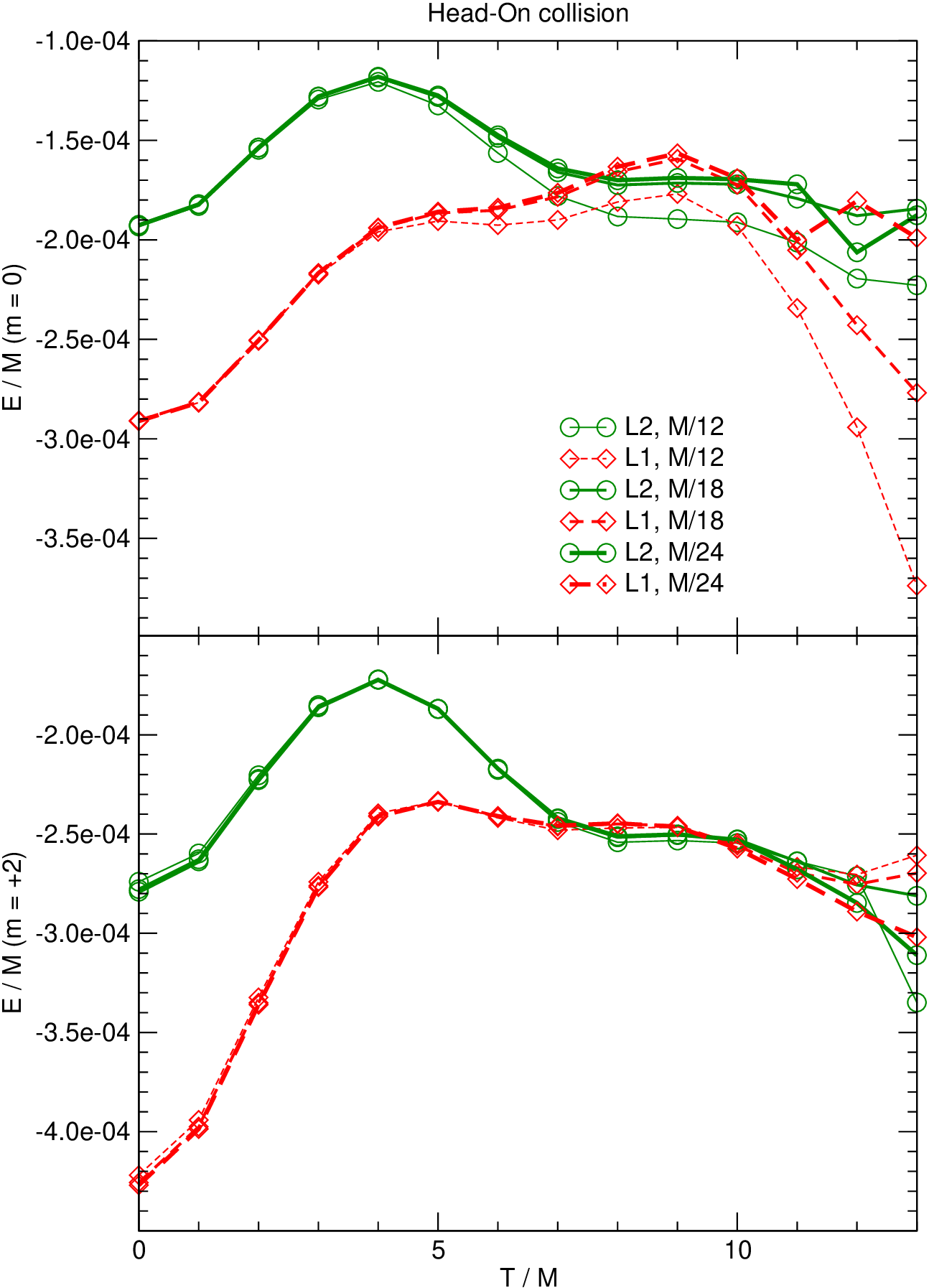}
  \end{center}
  \caption{Radiated energy for Head-On collision as calculated from
  original (L1) and new (L2) Lazarus methods, as a function of
  extraction time $T$. Note the extension of the plateau with
  increasing resolution.}
  \label{fig:HO-EvsT}
\end{figure}

Fig. \ref{fig:HO-EvsT} shows the total radiated energy from the
Teukolsky evolution, as a function of the extraction time $T$ for
original and new methods, for the $m = 0$ (upper panel) and $m = +2$
(lower panel) modes. For both sets of modes, original and new Lazarus
deviate at early extraction times. While the original Lazarus results
reproduce what was seen in Fig. 5 of \cite{Baker:2002qf} --- the
radiated energy reaches a crude plateau between extraction times of $4
M$ and $10 M$ --- the new-tetrad-produced energy reaches a level
plateau only after $7 \, M$. After this time, both methods agree well
until $10 \, M$, but with a noticeably more level plateau in the
weaker $m = 0$ curve for the new tetrad.

Table \ref{tab:HeadOn_emission} compares the \emph{total} energy
calculated with old and new Lazarus with a direct-extraction energy
taken from \cite{Zlochower:2005bj} (this extraction was from a rather
close detector position, $r_{*, obs} \approx 7.46\,M$). While both
Lazarus figures differ from the direct result, the new Lazarus plateau
-- measured from $T = 8\,M$ and $T = 11\,M$ -- has less than half the
standard deviation of old Lazarus, indicating a more stable plateau.

\begin{table}
  \caption{Comparison of emitted $E$ for Head-On collision. Direct
  results ($r_{*, obs} = 7.46\,M$) are from
  \cite{Zlochower:2005bj}. $\bar{E}$ and $\sigma_E$ are the mean and
  standard deviation from four plateau values.}
  \begin{ruledtabular}
    \begin{tabular}{r|rr}\label{tab:HeadOn_emission}
      Source & $|\bar{E}| / M$    & $\sigma_E / M$ \\
              & $ \times 10^{-4}$ & $\times 10^{-4}$  \\
      \hline
      Direct  & 6.6   &  ---  \\
      L1 ($8-11 M$) & 6.827 & 0.446 \\
      L2 ($8-11 M$) & 6.811 & 0.179 \\
    \end{tabular}

  \end{ruledtabular}
\end{table}

\begin{figure}
  \begin{center}
    \includegraphics*[width=3.3in]{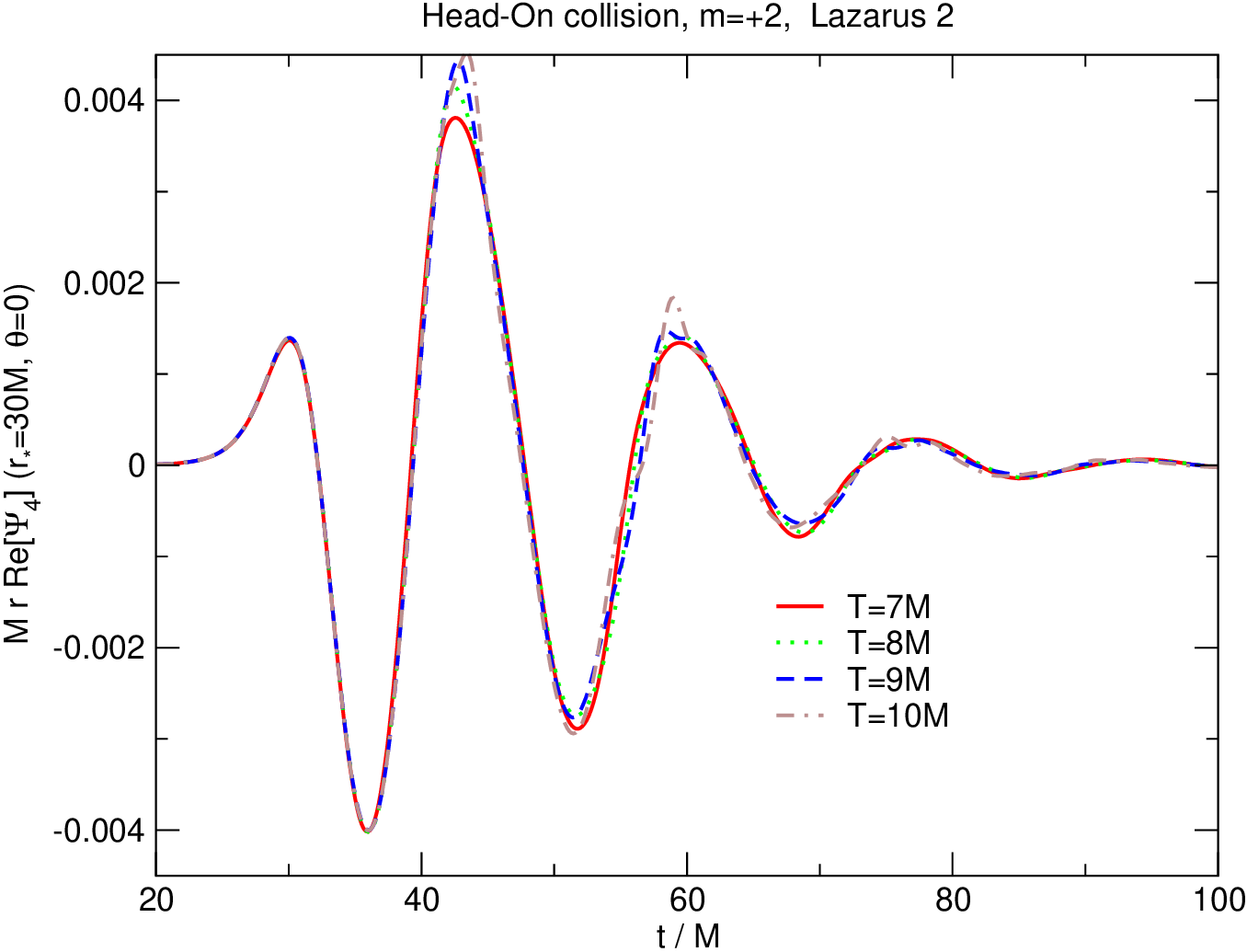}
  \end{center}
  \caption{Evolution of the $m = +2$ Head-On waveform for the new
  (L2) Lazarus method, evaluated at $(r_* = 30\,M, \theta = 0)$.}
  \label{fig:HO-WF-L2}
\end{figure}

Fig. \ref{fig:HO-WF-L2} shows the dominant-mode ($m = +2$) Lazarus
waveform for the Head-On collision, measured at $(r_* = 30\,M, \theta
= 0)$.  While the difference between curves for early extraction times
can give an idea of the uncertainties of the method when
nonlinearities are present the agreement at later times shows when a
safe, linear regime is reached.


\subsection{QC0 Data}

The final test of our method again revisits initial data treated in
\cite{Baker:2002qf}: Bowen-York binary black-hole data, where the
holes have zero spin, but are boosted in a direction transverse to
their separation, to achieve a net orbital angular momentum $J$.

Khanna et al. \cite{Khanna99a} investigated such data to first
perturbative order, through Zerilli and Teukolsky schemes. The emitted
energy curves from the two methods diverge at the level $J / M^2
\approx 0.4 - 0.5$, and this the authors take as the limit of
applicability of linear perturbative theory. At the highest analyzed
spin, $J / M^2 = 0.6$, the Teukolsky- and Zerilli-calculated energies
are $\sim 8 \times 10^{-3} \, M$ and $\sim 5 \times 10^{-3} \, M$
respectively.

Here, we use the binary Bowen-York data with equal bare masses $m =
0.45\,M$, located at coordinate positions $y = \pm 1.1515 \, M$, and
with Bowen-York ``boosts'' $P = \pm 0.335 \, M$ in the $x$
direction. These result in a physical throat-to-throat separation $L =
4.9 \, M$, and a total ADM angular momentum $J = 0.77 \, M^2$. We
refer to this data set as ``QC0'', following the classification scheme
of \cite{Baker:2002qf}; it was originally suggested by Baumgarte
\cite{Baumgarte00a} (adapting Cook's ``effective potential'' method
\cite{Cook94} to punctures) as a model for the so-called Innermost
Stable Circular Orbit (ISCO) of two equal-mass black holes.

In contrast to the two previous test cases, numerical instabilities
will kill the QC0 data evolution before a single common apparent
horizon has appeared (though a common event horizon may already be
present). Thus Lazarus extraction and evolution of Teukolsky data
would not appear to be completely justified at any time. Nevertheless,
\cite{Baker:2002qf} found a small plateau in the emitted energy,
indicating a range of extraction times where the system has
effectively linearized prior to the simulation's crash.

For this data, the Lazarus procedure was carried out initially using a
zeroth-order parameter estimate taken from the ADM energy and angular
momentum of the initial data: $M = 1$, $a = 0.77$. However, since the
emitted energy and angular momentum were at the level of a few percent
of the total, the subsequent drop in background mass and spin might be
significant. For this reason, we iterated the method with suitably
reduced background mass and spin parameters: $M = 0.974$, and $a =
0.675 / M = 0.693$.

\begin{figure}
  \begin{center}
    \includegraphics*[width=3.3in]{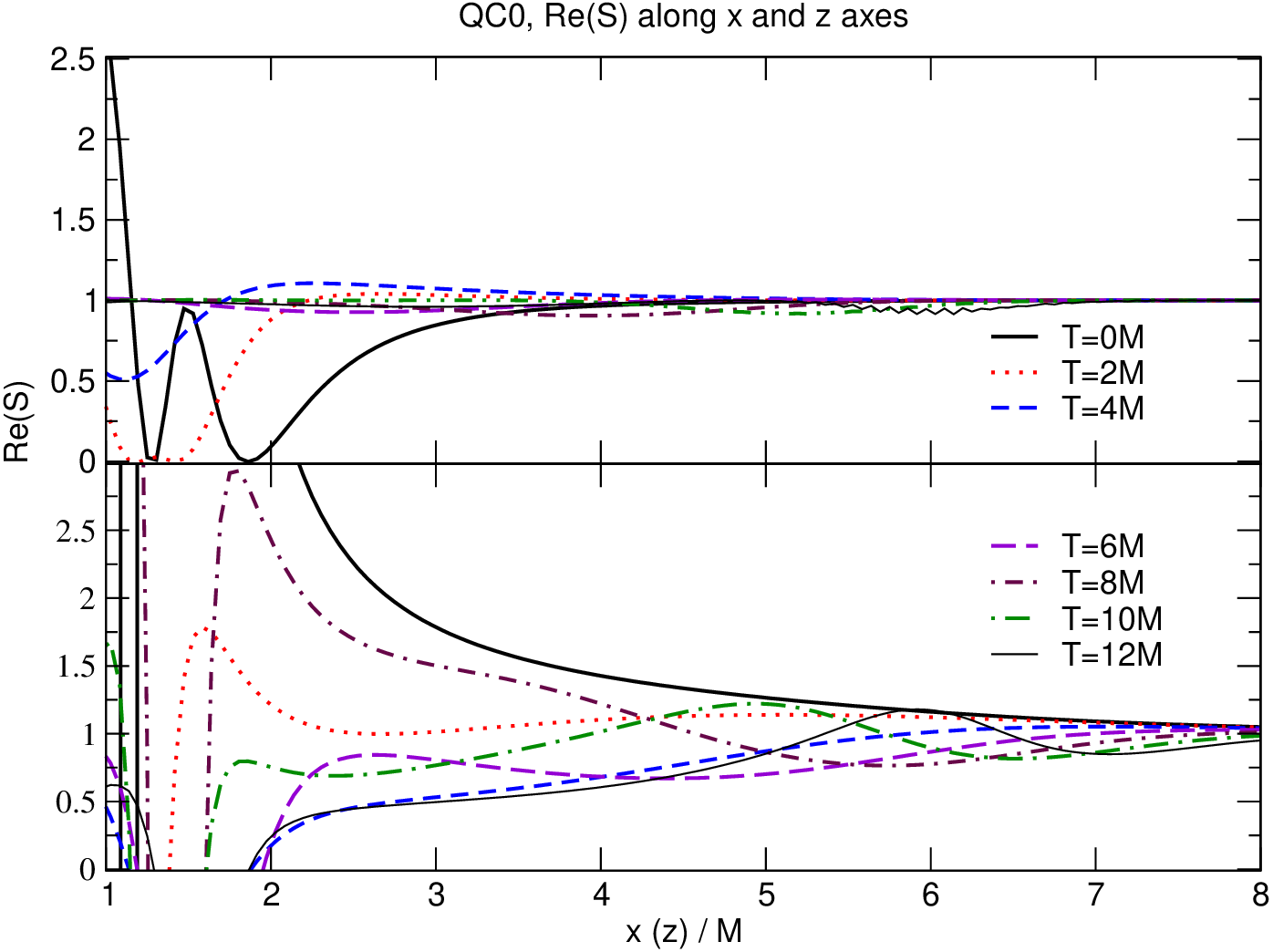}
  \end{center}
  \caption{The $\mathcal{S}$ invariant along the $x$ (upper) and $z$
    (lower) axes for QC0 data, at different extraction times $T$.}
  \label{fig:QC0-Sxz}
\end{figure}

To demonstrate how quickly the system appears to relax to Kerr, we
show in Fig. \ref{fig:QC0-Sxz} the $\mathcal{S}$ invariant along the
$x$ and $z$ axes. Along the $z$ axis, $\mathcal{S}$ clearly takes time
to settle down, only uniformly falling within $[0.5, 2.0]$ rather late
in the simulation, after $T = 8 \, M$.

\begin{figure}
  \begin{center}
    \includegraphics*[width=3.3in]{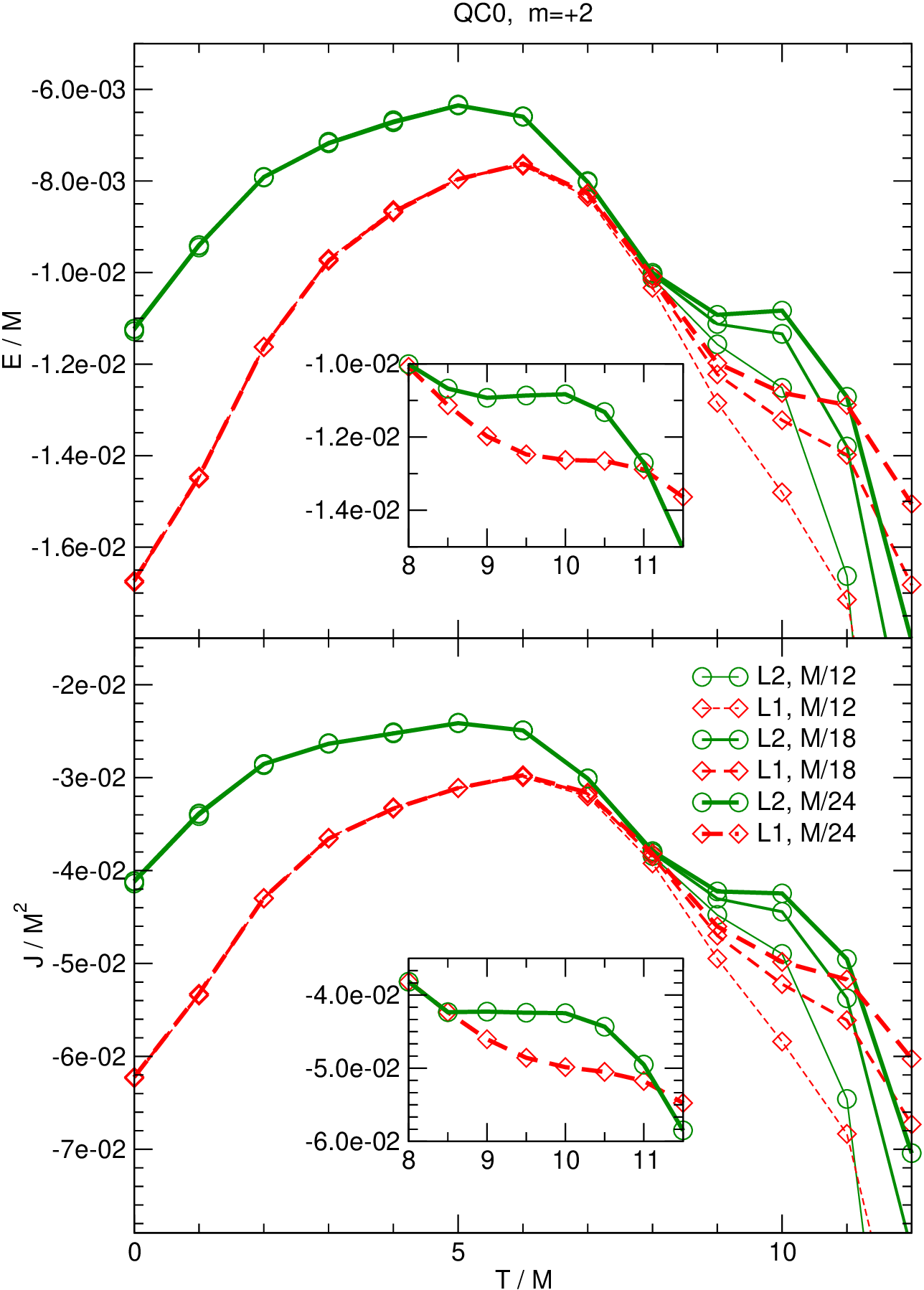}
  \end{center}
  \caption{$m=+2$ radiated energy and angular momentum for QC0 data
  as calculated from original (L1) and new (L2) Lazarus methods, as a
  function of extraction time $T$.}
  \label{fig:QC0-EJvsT}
\end{figure}

\begin{figure}
  \begin{center}
    \includegraphics*[width=3.3in]{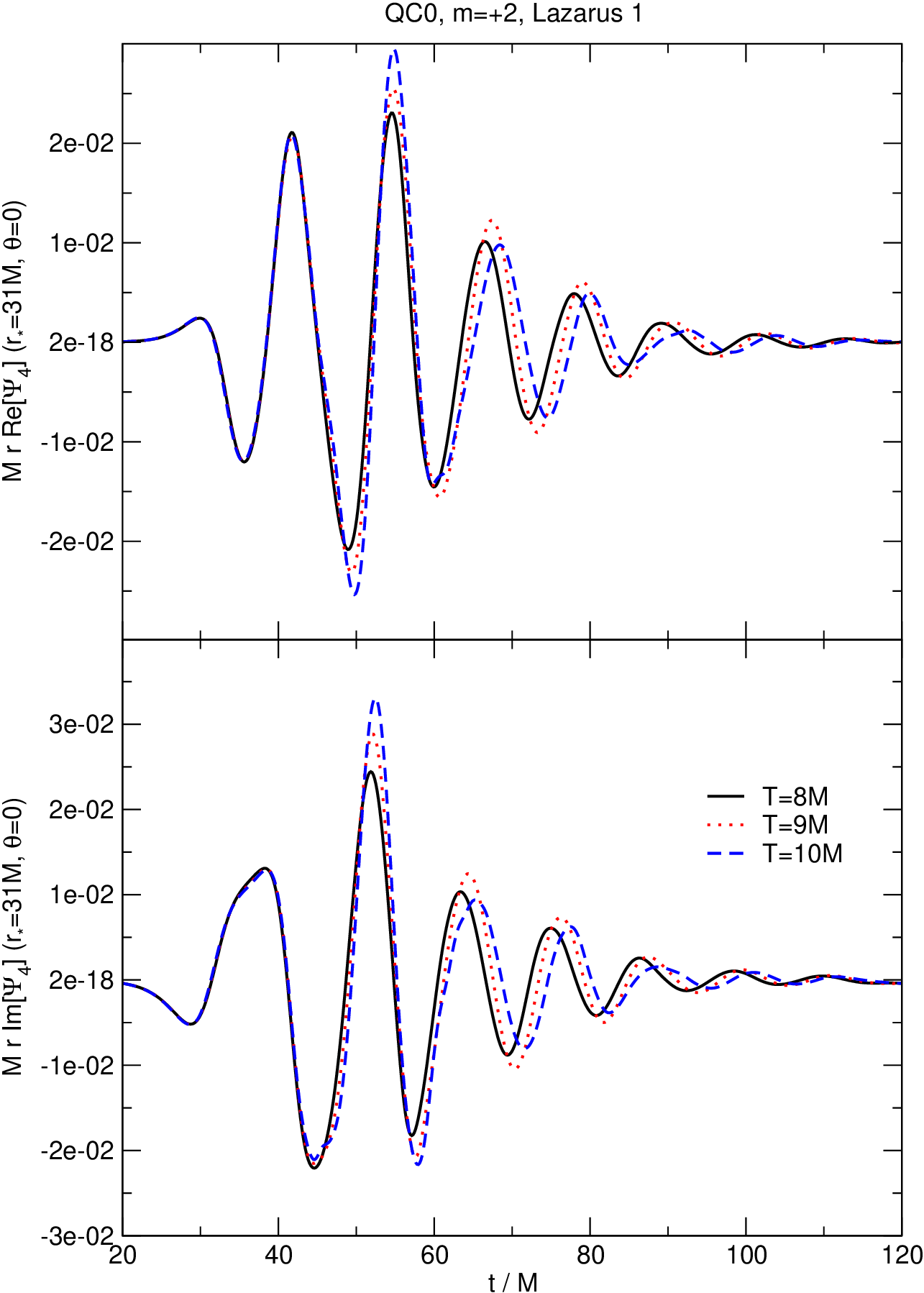}
  \end{center}
  \caption{The two polarizations of the $m = +2$ QC0 waveform for the
	original (L1) Lazarus method, evaluated at $(r_* = 31 \, M,
	\theta = 0)$.}
  \label{fig:QC0-WF-L1}
\end{figure}

\begin{figure}
  \begin{center}
    \includegraphics*[width=3.3in]{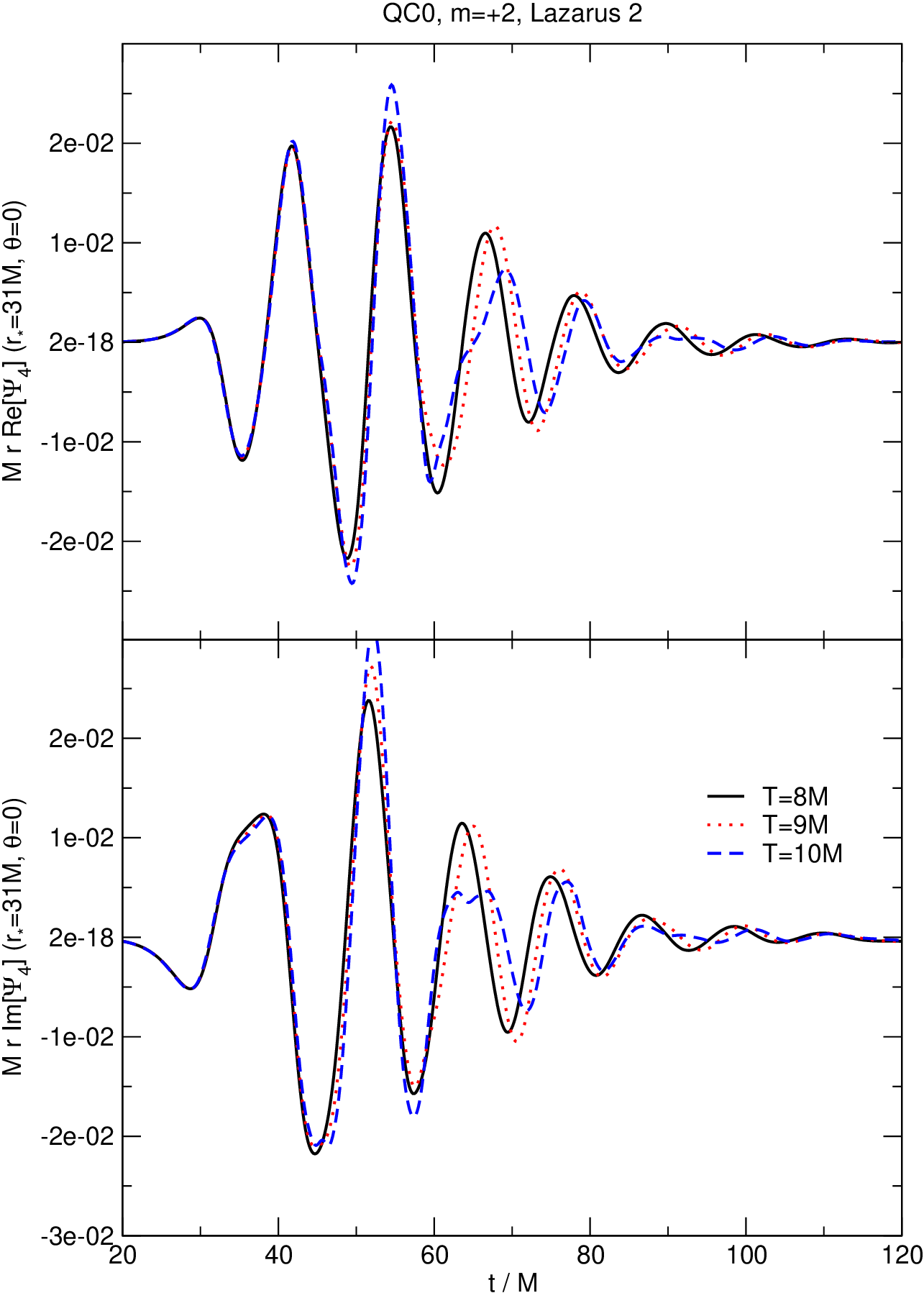}
  \end{center}
  \caption{The two polarizations of the $m = +2$ QC0 waveform for the
  new (L2) Lazarus method, evaluated at $(r_* = 31 \, M, \theta =
  0)$.}
  \label{fig:QC0-WF-L2}
\end{figure}

Fig. \ref{fig:QC0-EJvsT} shows the emitted energy and angular
momentum, respectively, for both original and new Lazarus methods (the
dominant $m = +2$ mode waveform is plotted for old and new Lazarus in
Fig. \ref{fig:QC0-WF-L1} and Fig. \ref{fig:QC0-WF-L2},
respectively). The original Lazarus results can be found in Fig. 22 of
\cite{Baker:2002qf} (note that the latter figure contains the total
energy summed across $m$ modes). There is significant deviation
between original and new Lazarus results up to the extraction time of
$T \approx 6\,M$. The two methods agree well for $T \in [7\, M, 8\,
M]$, but begin to diverge again around the time new Lazarus develops
plateau values of around $1.1 \times 10^{-2} \, M$ and $4.25 \times
10^{-2} \, M^2$ for the emitted energy and angular momentum, in the
mode $m=+2$, respectively. Working with the more time-resolved data of
the inserts in Fig. \ref{fig:QC0-EJvsT}, we have identified the
plateaus for old and new Lazarus, and calculated the means and
standard deviations. We present these in Table \ref{tab:QC0_emission},
along with direct-extraction results ($r_{*, obs} = 14.03\,M$) from
\cite{Campanelli:2005dd}. While both old and new means differ from the
direct-extraction figures, new Lazarus's plateau is flatter.

Because the QC0 data takes time to plunge to a single hole, the onset
of linearization will be delayed relative to, for instance, the
head-on case. {The fact that the observed plateau does not persist
for longer than $\sim 2 \,M$ is a consequence of the instability of
the the simple combination of ``ADM + maximal lapse + zero shift'' we
have used in our full-numerical evolutions here. In this sense, QC0 +
ADM is a ``marginal'' Lazarus case; to establish the plateau
unambiguously will require evolutions that last for several more
$M$. This will necessitate more advanced evolution systems and gauges,
and is addressed in the Discussion below.}

\begin{table}
  \caption{Comparison of emitted $E$ and $J$ for QC0 merger. Direct
  results ($r_{*, obs} = 14.03\,M$) are from
  \cite{Campanelli:2005dd}. $\bar{E}$, $\bar{J}$ are the means and
  $\sigma_E$, $\sigma_J$ the standard deviations from five plateau
  points (indicated in first column).}
  \begin{ruledtabular}
    \begin{tabular}{r|cc|cc}\label{tab:QC0_emission}
      Source & $|\bar{E}| / M$    & $\sigma_E / M$ & $|J| / M^2$ & $\sigma_J / M^2$ \\
              & $ \times 10^{-2}$ & $\times 10^{-2}$ & $ \times 10^{-2}$ & $ \times 10^{-2}$ \\
      \hline
      Direct            & $2.8 \pm 0.2$ &  ---  & $11.0 \pm 1.0$ & ---   \\
      L1 ($9.0-11.0 M$) & 2.570 & 0.067 &  9.493 & 0.434 \\
      L2 ($8.5-10.5 M$) & 2.422 & 0.031 &  8.553 & 0.175 \\
    \end{tabular}

  \end{ruledtabular}
\end{table}

\section{Discussion}\label{Sec:Discuss}

The original Lazarus method was a successful synthesis of
full-numerical and perturbative methods in numerical relativity. It
was responsible for the production of the first waveform for
black-hole binary mergers and close orbits, before these regimes were
accessible to a complete full-numerical treatment.

We have presented in this paper the first in a series of proposed
improvements to the Lazarus method --- improvements which will, we
hope, make Lazarus more ambitious in its reach, and more rigorous in
its grasp. The developments herein have focused on the important area
of tetrad determination. We can now obtain --- in the ``real time'' of
the full-numerical evolution --- a transverse tetrad that will differ
from the Kinnersley tetrad only by a spin-boost.

We described in Section \ref{Sec:Laz2Theory} how this new method of
tetrad determination can be incorporated into the Lazarus project,
eliminating the artificial mixing of monopolar, longitudinal and
transverse modes that was necessary in original Lazarus. To test this
new aspect, we have applied it on the same footing as original Lazarus
--- to the short-lived ADM evolutions of three different types of
black-hole data.

Our numerical results with this method, reported in \ref{Sec:Results},
have allowed us to test this mode mixing for the first time. The good
agreement between old and new methods for the single Spinning
Bowen-York and Head-On cases, where linearization is seen well before
code break-down, can be seen as a validation of the applicability of
the original Lazarus mixing formulas; additional evolution time may be
needed to resolve the marginal case of QC0 data satisfactorily. In
contrast, the areas of difference --- seen for instance, before $T = 7
\, M$ in Fig. \ref{fig:HO-EvsT} for the Head-On case --- indicate
better where our assumption of linear deviation from Kerr may not have
been justified. With this alternative path to the Kinnersley tetrad in
Lazarus, we have produced error estimates for the tetrad construction
procedure; we can estimate the contribution that errors in this
procedure make to the overall uncertainty in Lazarus energies. It
should be noted, however, that this error is unlikely to dominate the
\emph{total} uncertainty in emitted energies. Comparison of Lazarus
(both old and new) results with more direct extraction techniques,
shown in Tables \ref{tab:HeadOn_emission} and \ref{tab:QC0_emission},
indicate that other important factors must still be addressed. A
reasonable estimate of theoretical error in waveforms will be needed
for their application to analysis of gravitational-wave detector
signals in the future.

For the next stage in the Lazarus project, we plan to use modern
evolution schemes, such the one provided by the {\it LazEv} framework
\cite{Zlochower:2005bj}, which are more stable and accurate than ADM.
\emph{LazEv} currently supports higher-order finite differencing for
the BSSN formulation of Einstein's equations with the choice of
several dynamical gauge conditions, and should allow for extended
energy plateaus, extending from the time of first linearization of the
system until the time the radiation leaves the system. This added
full-numerical evolution time will aid in the unambiguous
identification of the linearization time for marginal cases such as
``QC0'' above.

At present, it is unclear how to relate Boyer-Lindquist time to the
numerical time developed with standard numerical gauges. Maximal
slicing was discussed in Section \ref{SSec:Laz1Results}; future
refinements of this may use analytic insights into the late-time shape
of the maximal lapse (see \cite{Reimann:2003zd} for work on maximal
slicings of Schwarzschild). Commonly used dynamic slicings (e.g.,
``1+log'' slicing) ensure a lapse function with the same qualitative
shape, falling smoothly off to unity at large distances. The exact
shape will vary greatly in the near-field region, however, and this
will affect both old and new Lazarus; this is an interesting problem,
which we hope to address in future work.

Treatment of the shift is decoupled from the lapse in the Lazarus
approach; the shift correction via adjustment of $\phi$ is independent
of the slicing analysis. Co-rotating coordinates can already be
accommodated with this method. We expect to treat more complicated
evolution shifts in a similar way.

The quasi-Kinnersley frame has further useful applications. For direct
radiation extraction at a finite observer location in the 3D numerical
grid. It also provides a closer interpretation in terms of radiation for
3-dimensional visualizations.
Originally one uses the numerical tetrad to evaluate
$\Psi_4$. In order to give this a direct interpretation in terms of radiation,
one implicitly assumes that the observer is sufficiently
far from the strong-field region that the background space is almost
flat, and the numerical tetrad is a good approximation to Kinnersley.

With the quasi-Kinnersley frame, we can get much closer to the true
Kinnersley form at finite observer locations. This should supply us
with a more robust waveform at distant locations, and allow us to
approach the near-field zone with greater confidence. While
impracticable for short-lived ADM evolutions, modern evolution systems
should produce several periods of directly extracted waveforms at
different extraction radii. Outer-radius waveforms may be compared
with Lazarus results; inner waveforms may be compared with outer ones
as a test of the quality of strong-field waveforms in general.

Aside from the longer full-numerical evolution times, we plan to
investigate further several issues to improve the generality of
Lazarus.

The transformations used to extract Boyer-Lindquist coordinates from
the numerical ones perform well, but there is much room for
improvement. One obvious --- and theoretically well-founded --- step
that can be taken is relaxing the assumption that there is a
one-to-one relationship between the numerical radial coordinate
$R_{\rm num}$ and the Boyer-Lindquist $r$. Since we know the
theoretical form of the invariant $\mathcal{I}$ as a function of $(r,
\theta)$, we can imagine using $\mathcal{I}$ everywhere to derive both
$r$ and $\theta$. We have performed initial investigations of this
possibility, but found that the off-equator radial transformation
derived does not reach sufficiently close to the horizon, at least for
the most difficult cases, such as the QC0. Longer full-numerical
evolutions may improve the performance of this technique, as a horizon
appears and circularizes. We also plan to revisit the choice of the
time slicing as this might be crucial to process longer-term fully
nonlinear evolutions.

\acknowledgments
The authors gratefully acknowledge the support of the NASA Center for
Gravitational Wave Astronomy at The University of Texas at Brownsville
(NAG5-13396), and NSF grants PHY-0140326 and PHY-0354867. All
numerical simulations were performed on the CGWA computer cluster {\it
Funes}.


\appendix

\section{Numerical-Kinnersley Transformations for Kerr Data}
\label{App_num2Kin}

In Boyer-Lindquist (BL) coordinates, the numerical tetrad --- defined
by (\ref{eqn:OT-null_tetrad}) with orthonormalized spherical coordinate
directions for the $\hat{e}_{(i)}$ --- takes the form:
\bea
\label{eqn:num_tetrad}
\vec{l}_{\rm num} &=& \frac{1}{2} \left[ \sqrt{\frac{\Omega}{\Delta \Sigma}} , \sqrt{\frac{\Delta}{\Sigma}} , 0, \frac{2 a M r}{\sqrt{\Delta \Omega \Sigma}} \right], \nonumber\\
\vec{n}_{\rm num} &=& \frac{1}{2} \left[ \sqrt{\frac{\Omega}{\Delta \Sigma}} , - \sqrt{\frac{\Delta}{\Sigma}} , 0, \frac{2 a M r}{\sqrt{\Delta \Omega \Sigma}} \right], \\
\vec{m}_{\rm num} &=& \frac{1}{2} \left[ 0 , 0, \frac{1}{\sqrt{\Sigma}} ,  \frac{i}{\sth} \, \sqrt{\frac{\Sigma}{\Omega}} \right]. \nonumber
\eea
Such a tetrad will differ strongly from the Kinnersley tetrad; as a
consequence, all Weyl scalars calculated from it will be non-zero. For
Kerr-BL, these values will be:
\bea
\label{eqn:Weyl_Kerr_num}
\Psi_0 = \Psi_4 & = & - \frac{M}{2 \Omega \bar{\zeta}^3} [3 ( \Lambda^2 - \Omega )]. \nonumber \\
\Psi_1 = - \Psi_3 & = & - \frac{M}{2 \Omega \bar{\zeta}^3} [3 i \Lambda \sqrt{\Lambda^2 - \Omega}]. \\
\Psi_2 & = & - \frac{M}{2 \Omega \bar{\zeta}^3} [- ( 3 \Lambda^2 - \Omega )]. \nonumber
\eea

For Kerr-BL coordinates, the numerical null tetrad
(\ref{eqn:num_tetrad}) used in 3+1 numerical calculations can be
transformed to the Kinnersley tetrad (\ref{eqn:Kin_tetrad}) via a
combination of null rotations and spin-boosts:
\bea
\label{eqn:num2Kin_tetrad}
\vec{l}_{\rm Kin} &=& \frac{F_A}{2} \left\{ (D + 1) \, \vec{l}_{\rm num} + (D - 1)
\, \vec{n}_{\rm num} \right. \nonumber\\
&& \left. - i \, A \, (\vec{m}_{\rm num} - \vec{\bar{m}}_{\rm num}) \right\},
\nonumber \\
\vec{n}_{\rm Kin} &=& \frac{F^{-1}_A}{2} \left\{ (D - 1) \, \vec{l}_{\rm num} + (D + 1)
\, \vec{n}_{\rm num} \right. \nonumber\\
&& \left. - i \, A \, (\vec{m}_{\rm num} - \vec{\bar{m}}_{\rm num}) \right\},\\
\vec{m}_{\rm Kin} &=& \frac{F_B}{2} \left\{ (D + 1) \, \vec{m}_{\rm num} - (D - 1)
\, \vec{\bar{m}}_{\rm num} \right. \nonumber\\
&& \left. + i \, A \, (\vec{l}_{\rm num} + \vec{n}_{\rm num}) \right\}. \nonumber
\eea
Note that this is a corrected form of the transformations that
appeared in Eqns (5.9a-c) of Paper I.

The dependence of the Kinnersley-tetrad $\Psi_4$ on the the
numerical-tetrad scalars was given in (\ref{eqn:num2Kin_Psi4}), for
perturbations of Kerr-BL coordinates. The following are the
corresponding expressions for the remaining Kinnersley Weyl scalars
(the mixing function $A$ and spin-boost functions $F_A$ and $F_B$ are
given by (\ref{eqn:params_num2Kin}), and $D \equiv \sqrt{A^2 + 1}$):
\bea
\Psi^{\rm Kin}_0 & = &  \left[ (D+1)^2 \, \Psi^{\rm num}_0 + 4 \, i \, A \, (D+1) \, \Psi^{\rm num}_1 \right. \nonumber \\
&& \left. - 6 \, A^2 \, \Psi^{\rm num}_2 - 4 \, i \, A \, (D-1) \, \Psi^{\rm num}_3 \right. \label{eqn:num2Kin_Psi0}\\
&& \left. + (D-1)^2 \, \Psi^{\rm num}_4 \right] F_A^2 \, F_B^2 / 4, \nonumber\\
\Psi^{\rm Kin}_1 & = & \left[ - i \, A \, (D + 1) \, \Psi^{\rm num}_0 + 2 \, (D + 1 + 2 \, A^2) \, \Psi^{\rm num}_1 \right. \nonumber \\
&& \left. + 6 \, i \, A \, D \, \Psi^{\rm num}_2 + 2 \, (D - 1 - 2 \, A^2) \, \Psi^{\rm num}_3 \right. \label{eqn:num2Kin_Psi1}\\
&& \left. - i \, A \, (D - 1) \, \Psi^{\rm num}_4 \right] F_A \, F_B / 4 , \nonumber\\
\Psi^{\rm Kin}_2 & = & \left[ A^2 \, \Psi^{\rm num}_0 - 4 \, i \, A \, D \, \Psi^{\rm num}_1 + (4 + 6 \, A^2) \, \Psi^{\rm num}_2 +\right. \nonumber \\
&& \left. 4 \, i \, A \, D \, \Psi^{\rm num}_3 - A^2 \, \Psi^{\rm num}_4 \right] / 4,  \label{eqn:num2Kin_Psi2}\\
\Psi^{\rm Kin}_3 & = & \left[ i \, A \, (D - 1) \, \Psi^{\rm num}_0 + 2 \, (D - 1 - 2 \, A^2) \, \Psi^{\rm num}_1 \right. \nonumber \\
&& \left. - 6 \, i \, A \, D \, \Psi^{\rm num}_2 + 2 \, (D + 1 + 2 \, A^2) \, \Psi^{\rm num}_3 \right. \label{eqn:num2Kin_Psi3} \\
&& \left. + i \, A \, (D + 1) \, \Psi^{\rm num}_4 \right] / (4 \, F_A \, F_B) . \nonumber
\eea
It can be easily verified that these transformations, applied to the
numerical-tetrad Weyl scalars of Kerr given in
(\ref{eqn:Weyl_Kerr_num}), produce the Kinnersley values:
\[ \Psi^{\rm Kin}_2 = M/\bar{\zeta}^3 \; , \; \Psi^{\rm Kin}_{i \neq 2} = 0. \]

\section{Perturbative Results}
\label{App_perturb_res}

\subsection{Weyl Scalars for Spinning Bowen-York Data}\label{App_BYSpin_perturb}

Gleiser et al. \cite{Gleiser:1998ng} investigated the low-spin
behavior of the Bowen-York data, obtaining an analytic solution for
the conformal factor accurate up to $O(\epsilon^3)$, where $\epsilon
\equiv J / M^2$, and evolving the extracted radiative modes via the
Zerilli formalism.

For zero spin, the Bowen-York solution reduces to the Schwarzschild
solution in isotropic coordinates, and is automatically
constraint-satisfying; for small spins, the dominant radiation ($\ell
= 2$) should scale as $\epsilon^2$, while the next mode ($\ell = 3$)
will scale as $\epsilon^3$.

As an estimate of the radiation content of this data, we may calculate
the Weyl scalars of the approximate solution above; using the
numerical tetrad (\ref{eqn:num_tetrad}), we find on the initial slice:
\bea
\label{eqn:Weyl_BYSpin_num}
\Psi_0 &=& \Psi_4 = \frac{1536}{5} \frac{M^3 \, \epsilon^2 \, R^5}{(2 \, R + M)^{14}} (A_0 + i \, \epsilon \, B_0) \ssth, \nonumber\\
\Psi_1 &=& - \Psi_3 = \frac{384}{5} \frac{M^2 \, \epsilon \, R^4}{(2 \, R + M)^{15}} (i \, A_1 + \epsilon \, B_1 + i \, \epsilon^2 \, C_1), \nonumber\\
\Psi_2 &=& \frac{64}{5} \frac{M \, R^3}{(2 \, R + M)^{14}} (A_2 + i \, \epsilon \, B_2 + \epsilon^2 \, C_2 \nonumber\\
       & & + i \, \epsilon^3 \, D_2),
\eea
where the $\epsilon$ coefficients in parentheses are:
\bean
A_0 &=& - (2 \, R + M)^2 (4 \, R^2 + 64 \, M \, R + M^2),\\
B_0 &=& - 96 \, M^2 \, R^2 \, \cth,\\
A_1 &=& - 5 \, (2 \, R - M) \, (2 \, R + M)^6,\\
B_1 &=& 16 \, M \, R \, (2 \, R + M)^4 \, (2 \, R - M) \, \cth, \\
C_1 &=& - 16 \, M^3 \, (2 \, R + M)^2 \, (14 \, R^2 - M \, R + M^2) \\
    & & + 672 \, M^3 \, R^3 \, (2 \, R - M) \, \ssth,\\
A_2 &=& 5 \, (2 \, R + M)^8,\\
B_2 &=& 60 \, M \, R (2 \, R + M)^6 \, \cth,\\
C_2 &=& 12 \, M^{12} \, (2 \, R + M)^2\\
    & & \times \left[ - (2 \, R + M)^2 \, (8 \, R^2 - 4 \, M \, R - M^2) \right.\\
    & & \left. + 4 \, R^2 \, (12 \, R^2 - 4 \, M \, R + 3 \, M^2) \, \ssth \right],\\
D_2 &=& 192 \, M^4 \, R \, \cth\\
    & & \times \left[ (2 \, R + M)^2 \, (4 \, R + M) - 36 \, R^3 \, \sin^3\theta \right].
\eean

For simplicity, we factor out $\kappa \equiv 64 \, M \, R^3 / (2 R +
M)^6$ from the $Q$ matrix; to $O(\epsilon)$, the three eigenvalues
of the reduced $Q$ matrix here are:
\bean
\mu_1 &=& 2 + i \, \epsilon \, \frac{24 \, M \, R \, \cth}{(2 \, R + M)^2},\\
\mu_2 &=& -1 - i \, \epsilon \, \frac{24 \, M \, R \, \cth}{(2 \, R + M)^2},\\
\mu_3 &=& - 1.
\eean
The desired eigenvalue is $\mu_1$, which tends to $-2 \times \mu_2$ or
$\mu_3$ as $\epsilon \rightarrow 0$. When the factor $\kappa$ is
put back in, the full eigenvalue is
\bean
\lambda_1 &=& 128 \, \frac{M \, R^3 [(2 \, R + M)^2 + 12 \, i \, \epsilon \, M \, R \, \cth]}{(2 \, R + M)^8}\\
          &=& 2 \, \frac{M \, (r + 3 \, i \, a \, \cth)}{r^4},
\eean
where we've interpreted $a \equiv M \epsilon$ as the (dimensional)
Kerr spin parameter, reintroduced the Kerr-BL radial variable $r
\equiv R + M + (M^2 - a^2)/(4 R)$, and neglected terms of higher than
linear order. The corresponding eigenvector will be (any multiple of):
\[
\vec{V} = \left[ \frac{4 \, i \, M \, R \, \sth \, (2 \, R - M )}{(2 \, R + M)^3} \epsilon , 0, 1 \right]
\]
Note that $\vec{V}$ becomes real at $\theta = 0$, and
at $R = M / 2$.

\subsection{Weyl Scalars for Head-On Data}

The data used for the head-on run was of the Brill-Lindquist type,
with holes separated along the $y$ axis.

As an estimate of the radiation content of this data, we may calculate
the Weyl scalars of the approximate solution above; using the
numerical tetrad (\ref{eqn:num_tetrad}), we find on the initial slice:
\bea
\Psi_0 = \bar{\Psi}_4 &=& - 48 \, \epsilon^2 \, M^3 \, \frac{(\ccph - \ccth \, \ssph)}{(2 \, R + M)^5} \nonumber\\
&& + 96 \, i \, \epsilon^2 \, M^3 \, \frac{ \cth \, \sph \, \cph}{(2 \, R + M)^5}, \nonumber\\
\Psi_1 = - \bar{\Psi}_3 &=& 48 \, \epsilon^2 \, M^3 \, \frac{(8 \, R + M) \, \sth \, \cth \, \ssph}{(2 \, R + M)^6} \nonumber\\
&& + 48 \, i \, \epsilon^2 \, M^3 \, \frac{(8 \, R + M) \, \sth \, \sph \, \cph}{(2 \, R + M)^6} \nonumber\\
\Psi_2 &=& \frac{64 \, M \, R^3}{(2 R + M)^6}+ 16 \, \epsilon^2 \, \frac{M^3}{(2 R + M)^7}\\
&& \times (24 \, R^2 + 2 \, M \, R + M^2) \, T(\theta,\phi) \nonumber,
\eea
where we define $T(\theta,\phi) \equiv 2 - 3 \, (\ccth + \ssth \,
\ccph)$. For simplicity, we factor out $\kappa \equiv 64 \, M \, R^3 /
(2 R + M)^6$ from the $Q$ matrix; the three eigenvalues of the reduced
$Q$ matrix are:
\bean
\mu_1 &=& - 1,\\
\mu_2 &=& 2 + \epsilon^2 \, \frac{M^2 \, (24 \, R^2 + 2 \, M \, R + M^2) \, T(\theta,\phi)}{2 \, R^3 \, (2 \, R + M)},\\
\mu_3 &=& -1 - \epsilon^2 \, \frac{M^2 \, (24 \, R^2 + 2 \, M \, R + M^2) \, T(\theta,\phi)}{2 \, R^3 \, (2 \, R + M)}.
\eean
Here the principal eigenvalue is obviously $\mu_2$, and the
rescaled equivalent is:
\bean
\lambda_2 &=& \frac{128 \, M \, R^3}{(2 \, R + M)^6} \\
&& + \epsilon^2 \, \frac{32 \,M^3 \, R^3 \, (24 \, R^2 + 2 \, M \, R + M^2) \, T(\theta,\phi)}{R^3 \, (2 \, R + M)^7}\\
&=& \frac{128 \, M \, R^3}{(2 \, R + M)^6} \\
&& + \frac{32 \, d^2 \, M \, R^3 \, (24 \, R^2 + 2 \, M \, R + M^2) \, T(\theta,\phi)}{R^3 \, (2 \, R + M)^7}\\
\eean
with a related normalized (to $O(\epsilon^2)$) eigenvector
\bean
\vec{V} &=& \left[ \begin{array}{ccc} - \epsilon^2 \, M^2 \, (8 \, R + M) \, \sth \, \cth \, \ssph / 2 \, R^3 \\ \epsilon^2 \, M^2 \, (8 \, R + M) \, \sth \, \sph \, \cph / 2 \, R^3 \\ 1 \end{array} \right]\\
&=& \left[ \begin{array}{ccc} - d^2 \, (8 \, R + M) \, \sth \, \cth \, \ssph / 2 \, R^3 \\ d^2 \, (8 \, R + M) \, \sth \, \sph \, \cph / 2 \, R^3 \\ 1 \end{array} \right]
\eean

Note that in this case, the normalized eigenvector is manifestly real
everywhere, and so would lead to a degeneracy problem in the
reconstruction of the quasi-Kinnersley tetrad, of the type described
in Section \ref{SSec:NewTetrad}.

\subsection{Weyl Scalars for QC0 Data}\label{App_QC0_perturb}

The data used for the QC0 run was of the binary Bowen-York type. We
can try to determine some properties of the slow-close limiting form
of this data. This was addressed by \cite{Khanna99a}, who treated the
binary data --- with zero-spin holes separated by $L$ in coordinate
space, and transversely boosted with momenta $\pm P$ --- as a
perturbation of Schwarzschild with perturbation parameter $\epsilon
\equiv L \, P / M^2$. The authors worked solely in the Zerilli
formalism (they point out that such data is only a perturbation of
Schwarzschild, not of Kerr), and only to $O(\epsilon)$, where the
Hamiltonian constraint did not need to be solved for consistency.

Taking this data (separation along the $y$ axis, and boost in the
$x$ direction), the numerical tetrad yields
\bean
\Psi_2 &=& \frac{64 \, M \, R^3}{(2 R + M)^8} \left[ (2 R + M)^2 - 12 \, i \epsilon \, M \, R \, \cth \right],\\
\Psi_1 &=& \bar{\Psi}_3 = - \frac{384 \, \epsilon \, M^2 \, R^4 \, \sth}{(2 R + M)^9} \left[ 4 (2 R + M) \cos\phi \, \sin\phi \, \cth \right.\\
       && \left. - 6 \, i \, R - i \, M + 4 \, i (2 R + M) \cos^2\phi \right],\\
\Psi_0 &=& - \Psi_4 = - \frac{768 \, \epsilon \, M^2 \, R^4 (2 R - M)}{(2 R + M)^9} \\
       && \times \left[ \cos\phi \, \sin\phi (1 + \ccth) - i \cth (1 - 2 \cos^2\phi) \right].
\eean
For simplicity, we factor out $\kappa \equiv 64 \, M \, R^3 / (2 R +
M)^6$ from the $Q$ matrix; the three eigenvalues of the reduced $Q$
matrix are:
\bean
\mu_1 &=& 2 - i \, \epsilon \, \frac{24 \, M \, R \, \cth}{(2 R + M)^2},\\
\mu_2 &=& -1 + i \, \epsilon \, \frac{24 \, M \, R \, \cth}{(2 R + M)^2},\\
\mu_3 &=& - 1.
\eean
The desired eigenvalue is $\mu_1$, which tends to $-2 \times \mu_2$ or
$\mu_3$ as $\epsilon \rightarrow 0$. When the factor $\kappa$ is put
back in, the full eigenvalue is
\bean
\lambda_1 &=& 128 \frac{M R^3 [(2 R + M)^2 - 12 i \, \epsilon \, M \, R \, \cth]}{(2 R + M)^8}\\
          &=& 2 \frac{M (r + 3 i a \cth)}{r^4},
\eean
where we've interpreted $a \equiv - M \epsilon$ as the (dimensional)
Kerr spin parameter, reintroduced the Kerr-BL radial variable $r
\equiv R + M + (M^2 - a^2)/(4 R)$, and neglected terms of higher than
linear order. The corresponding eigenvector will be (any multiple of):
\bean
\vec{V} &=& \left[ \frac{4 \, i \, M \, R \, \sth \, (-6 \, R - M + 4 (2 R + M)\, \cos^2\phi )}{(2 R + M)^3} \epsilon , \right.\\
        && \left. \frac{16 \, i \, M \, R \, \sth \, \cth \, \sin\phi \, \cos\phi )}{(2 R + M)^2} \epsilon , 1 \right]
\eean
Note that $\vec{V}$ becomes real for certain angular positions: e.g.,
$\theta = 0$ for all $\phi$, and $\theta = \pi / 2$ for $\phi =
\arccos \left( \sqrt{\frac{6 \, R + M}{4 ( 2 \, R + M)}}\right)
\rightarrow \pi / 6$ for $R >> M$.

%


\bibliographystyle{apsrev}

\begin{thebibliography}{74}
\expandafter\ifx\csname natexlab\endcsname\relax\def\natexlab#1{#1}\fi
\expandafter\ifx\csname bibnamefont\endcsname\relax
  \def\bibnamefont#1{#1}\fi
\expandafter\ifx\csname bibfnamefont\endcsname\relax
  \def\bibfnamefont#1{#1}\fi
\expandafter\ifx\csname citenamefont\endcsname\relax
  \def\citenamefont#1{#1}\fi
\expandafter\ifx\csname url\endcsname\relax
  \def\url#1{\texttt{#1}}\fi
\expandafter\ifx\csname urlprefix\endcsname\relax\def\urlprefix{URL }\fi
\providecommand{\bibinfo}[2]{#2}
\providecommand{\eprint}[2][]{\url{#2}}

\bibitem[{\citenamefont{Price and Pullin}(1994)}]{Price94a}
\bibinfo{author}{\bibfnamefont{R.~H.} \bibnamefont{Price}} \bibnamefont{and}
  \bibinfo{author}{\bibfnamefont{J.}~\bibnamefont{Pullin}},
  \bibinfo{journal}{Phys. Rev. Lett.} \textbf{\bibinfo{volume}{72}},
  \bibinfo{pages}{3297} (\bibinfo{year}{1994}).

\bibitem[{\citenamefont{Pullin}(1998)}]{Pullin98a}
\bibinfo{author}{\bibfnamefont{J.}~\bibnamefont{Pullin}}, in
  \emph{\bibinfo{booktitle}{Proceedings of GR15}}, edited by
  \bibinfo{editor}{\bibfnamefont{N.}~\bibnamefont{Dadhich}} \bibnamefont{and}
  \bibinfo{editor}{\bibfnamefont{J.}~\bibnamefont{Narlikar}}
  (\bibinfo{publisher}{Inter-Univ. Centre for Astron. and Astrophys.},
  \bibinfo{address}{Pune, India}, \bibinfo{year}{1998}),
  p.~\bibinfo{pages}{87}, \eprint{gr-qc/9803005}.

\bibitem[{\citenamefont{Lousto}(2001)}]{Lousto99a}
\bibinfo{author}{\bibfnamefont{C.~O.} \bibnamefont{Lousto}},
  \bibinfo{journal}{Phys. Rev. D} \textbf{\bibinfo{volume}{63}},
  \bibinfo{pages}{047504} (\bibinfo{year}{2001}), \eprint{gr-qc/9911109}.

\bibitem[{\citenamefont{Hahn and Lindquist}(1964)}]{Hahn64}
\bibinfo{author}{\bibfnamefont{S.~G.} \bibnamefont{Hahn}} \bibnamefont{and}
  \bibinfo{author}{\bibfnamefont{R.~W.} \bibnamefont{Lindquist}},
  \bibinfo{journal}{Ann. Phys.} \textbf{\bibinfo{volume}{29}},
  \bibinfo{pages}{304} (\bibinfo{year}{1964}).

\bibitem[{\citenamefont{Smarr}(1975)}]{Smarr75}
\bibinfo{author}{\bibfnamefont{L.}~\bibnamefont{Smarr}}, Ph.D. thesis,
  \bibinfo{school}{University of Texas, Austin}, \bibinfo{address}{Austin,
  Texas} (\bibinfo{year}{1975}).

\bibitem[{\citenamefont{Smarr et~al.}(1976)\citenamefont{Smarr, \v{C}ade\v{z},
  DeWitt, and Eppley}}]{Smarr76}
\bibinfo{author}{\bibfnamefont{L.}~\bibnamefont{Smarr}},
  \bibinfo{author}{\bibfnamefont{A.}~\bibnamefont{\v{C}ade\v{z}}},
  \bibinfo{author}{\bibfnamefont{B.}~\bibnamefont{DeWitt}}, \bibnamefont{and}
  \bibinfo{author}{\bibfnamefont{K.}~\bibnamefont{Eppley}},
  \bibinfo{journal}{Phys. Rev. D} \textbf{\bibinfo{volume}{14}},
  \bibinfo{pages}{2443} (\bibinfo{year}{1976}).

\bibitem[{\citenamefont{\v{C}ade\v{z}}(1971)}]{Cadez71}
\bibinfo{author}{\bibfnamefont{A.}~\bibnamefont{\v{C}ade\v{z}}}, Ph.D. thesis,
  \bibinfo{school}{University of North Carolina at Chapel Hill},
  \bibinfo{address}{Chapel Hill, North Carolina} (\bibinfo{year}{1971}).

\bibitem[{\citenamefont{\v{C}ade\v{z}}(1974)}]{Cadez74}
\bibinfo{author}{\bibfnamefont{A.}~\bibnamefont{\v{C}ade\v{z}}},
  \bibinfo{journal}{Ann. Phys.} \textbf{\bibinfo{volume}{83}},
  \bibinfo{pages}{449} (\bibinfo{year}{1974}).

\bibitem[{\citenamefont{Anninos and Brandt}(1998)}]{Anninos98a}
\bibinfo{author}{\bibfnamefont{P.}~\bibnamefont{Anninos}} \bibnamefont{and}
  \bibinfo{author}{\bibfnamefont{S.}~\bibnamefont{Brandt}},
  \bibinfo{journal}{Phys. Rev. Lett.} \textbf{\bibinfo{volume}{81}},
  \bibinfo{pages}{508} (\bibinfo{year}{1998}), \eprint{gr-qc/9806031}.

\bibitem[{\citenamefont{Baker et~al.}(2000)\citenamefont{Baker, Br\"ugmann,
  Campanelli, and Lousto}}]{Baker00b}
\bibinfo{author}{\bibfnamefont{J.}~\bibnamefont{Baker}},
  \bibinfo{author}{\bibfnamefont{B.}~\bibnamefont{Br\"ugmann}},
  \bibinfo{author}{\bibfnamefont{M.}~\bibnamefont{Campanelli}},
  \bibnamefont{and} \bibinfo{author}{\bibfnamefont{C.~O.}
  \bibnamefont{Lousto}}, \bibinfo{journal}{Class. Quantum Grav.}
  \textbf{\bibinfo{volume}{17}}, \bibinfo{pages}{L149} (\bibinfo{year}{2000}),
  \eprint{gr-qc/0003027}.

\bibitem[{\citenamefont{Baker et~al.}(2002{\natexlab{a}})\citenamefont{Baker,
  Campanelli, and Lousto}}]{Baker:2001sf}
\bibinfo{author}{\bibfnamefont{J.}~\bibnamefont{Baker}},
  \bibinfo{author}{\bibfnamefont{M.}~\bibnamefont{Campanelli}},
  \bibnamefont{and} \bibinfo{author}{\bibfnamefont{C.~O.}
  \bibnamefont{Lousto}}, \bibinfo{journal}{Phys. Rev. D}
  \textbf{\bibinfo{volume}{65}}, \bibinfo{pages}{044001}
  (\bibinfo{year}{2002}{\natexlab{a}}), \eprint{gr-qc/0104063 (misprints
  corrected, 2005)}.

\bibitem[{\citenamefont{Baker et~al.}(2001)\citenamefont{Baker, Br{\"u}gmann,
  Campanelli, Lousto, and Takahashi}}]{Baker:2001nu}
\bibinfo{author}{\bibfnamefont{J.}~\bibnamefont{Baker}},
  \bibinfo{author}{\bibfnamefont{B.}~\bibnamefont{Br{\"u}gmann}},
  \bibinfo{author}{\bibfnamefont{M.}~\bibnamefont{Campanelli}},
  \bibinfo{author}{\bibfnamefont{C.~O.} \bibnamefont{Lousto}},
  \bibnamefont{and}
  \bibinfo{author}{\bibfnamefont{R.}~\bibnamefont{Takahashi}},
  \bibinfo{journal}{Phys. Rev. Lett.} \textbf{\bibinfo{volume}{87}},
  \bibinfo{pages}{121103} (\bibinfo{year}{2001}),
  \eprint[http://arXiv.org/abs]{gr-qc/0102037}.

\bibitem[{\citenamefont{Baker et~al.}(2002{\natexlab{b}})\citenamefont{Baker,
  Campanelli, Lousto, and Takahashi}}]{Baker:2002qf}
\bibinfo{author}{\bibfnamefont{J.}~\bibnamefont{Baker}},
  \bibinfo{author}{\bibfnamefont{M.}~\bibnamefont{Campanelli}},
  \bibinfo{author}{\bibfnamefont{C.~O.} \bibnamefont{Lousto}},
  \bibnamefont{and}
  \bibinfo{author}{\bibfnamefont{R.}~\bibnamefont{Takahashi}},
  \bibinfo{journal}{Phys. Rev. D} \textbf{\bibinfo{volume}{65}},
  \bibinfo{pages}{124012} (\bibinfo{year}{2002}{\natexlab{b}}),
  \eprint[http://arXiv.org/abs]{astro-ph/0202469}.

\bibitem[{\citenamefont{Baker et~al.}(2004)\citenamefont{Baker, Campanelli,
  Lousto, and Takahashi}}]{Baker:2004wv}
\bibinfo{author}{\bibfnamefont{J.}~\bibnamefont{Baker}},
  \bibinfo{author}{\bibfnamefont{M.}~\bibnamefont{Campanelli}},
  \bibinfo{author}{\bibfnamefont{C.~O.} \bibnamefont{Lousto}},
  \bibnamefont{and}
  \bibinfo{author}{\bibfnamefont{R.}~\bibnamefont{Takahashi}},
  \bibinfo{journal}{Phys. Rev. D} \textbf{\bibinfo{volume}{69}},
  \bibinfo{pages}{027505} (\bibinfo{year}{2004}).

\bibitem[{\citenamefont{Campanelli}(2005)}]{Campanelli:2004zw}
\bibinfo{author}{\bibfnamefont{M.}~\bibnamefont{Campanelli}},
  \bibinfo{journal}{Class. Quant. Grav.} \textbf{\bibinfo{volume}{22}},
  \bibinfo{pages}{S387} (\bibinfo{year}{2005}), \eprint{astro-ph/0411744}.

\bibitem[{\citenamefont{Nakamura et~al.}(1987)\citenamefont{Nakamura, Oohara,
  and Kojima}}]{Nakamura87}
\bibinfo{author}{\bibfnamefont{T.}~\bibnamefont{Nakamura}},
  \bibinfo{author}{\bibfnamefont{K.}~\bibnamefont{Oohara}}, \bibnamefont{and}
  \bibinfo{author}{\bibfnamefont{Y.}~\bibnamefont{Kojima}},
  \bibinfo{journal}{Prog. Theor. Phys. Suppl.} \textbf{\bibinfo{volume}{90}},
  \bibinfo{pages}{1} (\bibinfo{year}{1987}).

\bibitem[{\citenamefont{Shibata and Nakamura}(1995)}]{Shibata95}
\bibinfo{author}{\bibfnamefont{M.}~\bibnamefont{Shibata}} \bibnamefont{and}
  \bibinfo{author}{\bibfnamefont{T.}~\bibnamefont{Nakamura}},
  \bibinfo{journal}{Phys. Rev. D} \textbf{\bibinfo{volume}{52}},
  \bibinfo{pages}{5428} (\bibinfo{year}{1995}).

\bibitem[{\citenamefont{Friedrich}(1996)}]{Friedrich96}
\bibinfo{author}{\bibfnamefont{H.}~\bibnamefont{Friedrich}},
  \bibinfo{journal}{Class. Quantum Grav.} \textbf{\bibinfo{volume}{13}},
  \bibinfo{pages}{1451} (\bibinfo{year}{1996}).

\bibitem[{\citenamefont{Frittelli and Reula}(1996)}]{Frittelli:1996wr}
\bibinfo{author}{\bibfnamefont{S.}~\bibnamefont{Frittelli}} \bibnamefont{and}
  \bibinfo{author}{\bibfnamefont{O.~A.} \bibnamefont{Reula}},
  \bibinfo{journal}{Phys. Rev. Lett.} \textbf{\bibinfo{volume}{76}},
  \bibinfo{pages}{4667} (\bibinfo{year}{1996}), \eprint{gr-qc/9605005}.

\bibitem[{\citenamefont{Baumgarte and Shapiro}(1999)}]{Baumgarte99}
\bibinfo{author}{\bibfnamefont{T.~W.} \bibnamefont{Baumgarte}}
  \bibnamefont{and} \bibinfo{author}{\bibfnamefont{S.~L.}
  \bibnamefont{Shapiro}}, \bibinfo{journal}{Phys. Rev. D}
  \textbf{\bibinfo{volume}{59}}, \bibinfo{pages}{024007}
  (\bibinfo{year}{1999}), \eprint{gr-qc/9810065}.

\bibitem[{\citenamefont{Anderson and York}(1999)}]{Anderson99}
\bibinfo{author}{\bibfnamefont{A.}~\bibnamefont{Anderson}} \bibnamefont{and}
  \bibinfo{author}{\bibfnamefont{J.~W.} \bibnamefont{York}},
  \bibinfo{journal}{Phy. Rev. Lett.} \textbf{\bibinfo{volume}{82}},
  \bibinfo{pages}{4384} (\bibinfo{year}{1999}), \eprint{gr-qc/9901021}.

\bibitem[{\citenamefont{Alcubierre et~al.}(2001)\citenamefont{Alcubierre,
  Benger, Br\"ugmann, Lanfermann, Nerger, Seidel, and
  Takahashi}}]{Alcubierre00b}
\bibinfo{author}{\bibfnamefont{M.}~\bibnamefont{Alcubierre}},
  \bibinfo{author}{\bibfnamefont{W.}~\bibnamefont{Benger}},
  \bibinfo{author}{\bibfnamefont{B.}~\bibnamefont{Br\"ugmann}},
  \bibinfo{author}{\bibfnamefont{G.}~\bibnamefont{Lanfermann}},
  \bibinfo{author}{\bibfnamefont{L.}~\bibnamefont{Nerger}},
  \bibinfo{author}{\bibfnamefont{E.}~\bibnamefont{Seidel}}, \bibnamefont{and}
  \bibinfo{author}{\bibfnamefont{R.}~\bibnamefont{Takahashi}},
  \bibinfo{journal}{Phys. Rev. Lett.} \textbf{\bibinfo{volume}{87}},
  \bibinfo{pages}{271103} (\bibinfo{year}{2001}), \eprint{gr-qc/0012079}.

\bibitem[{\citenamefont{Kidder et~al.}(2001)\citenamefont{Kidder, Scheel, and
  {T}eukolsky}}]{Kidder01a}
\bibinfo{author}{\bibfnamefont{L.~E.} \bibnamefont{Kidder}},
  \bibinfo{author}{\bibfnamefont{M.~A.} \bibnamefont{Scheel}},
  \bibnamefont{and} \bibinfo{author}{\bibfnamefont{S.~A.}
  \bibnamefont{{T}eukolsky}}, \bibinfo{journal}{Phys. Rev. D}
  \textbf{\bibinfo{volume}{64}}, \bibinfo{pages}{064017}
  (\bibinfo{year}{2001}), \eprint{gr-qc/0105031}.

\bibitem[{\citenamefont{Sarbach and Tiglio}(2002)}]{Sarbach02b}
\bibinfo{author}{\bibfnamefont{O.}~\bibnamefont{Sarbach}} \bibnamefont{and}
  \bibinfo{author}{\bibfnamefont{M.}~\bibnamefont{Tiglio}},
  \bibinfo{journal}{Phys. Rev. D} \textbf{\bibinfo{volume}{66}},
  \bibinfo{pages}{064023} (\bibinfo{year}{2002}).

\bibitem[{\citenamefont{Shinkai and Yoneda}(2002)}]{Shinkai02a}
\bibinfo{author}{\bibfnamefont{H.}~\bibnamefont{Shinkai}} \bibnamefont{and}
  \bibinfo{author}{\bibfnamefont{G.}~\bibnamefont{Yoneda}}
  (\bibinfo{year}{2002}), \eprint{gr-qc/0209111}.

\bibitem[{\citenamefont{Lindblom and Scheel}(2003)}]{Lindblom:2003ad}
\bibinfo{author}{\bibfnamefont{L.}~\bibnamefont{Lindblom}} \bibnamefont{and}
  \bibinfo{author}{\bibfnamefont{M.~A.} \bibnamefont{Scheel}},
  \bibinfo{journal}{Phys. Rev. D} \textbf{\bibinfo{volume}{67}},
  \bibinfo{pages}{124005} (\bibinfo{year}{2003}), \eprint{gr-qc/0301120}.

\bibitem[{\citenamefont{Bona and Palenzuela}(2004)}]{Bona:2004yp}
\bibinfo{author}{\bibfnamefont{C.}~\bibnamefont{Bona}} \bibnamefont{and}
  \bibinfo{author}{\bibfnamefont{C.}~\bibnamefont{Palenzuela}},
  \bibinfo{journal}{Phys. Rev. D} \textbf{\bibinfo{volume}{69}},
  \bibinfo{pages}{104003} (\bibinfo{year}{2004}), \eprint{gr-qc/0401019}.

\bibitem[{\citenamefont{Pretorius}(2005{\natexlab{a}})}]{Pretorius:2004jg}
\bibinfo{author}{\bibfnamefont{F.}~\bibnamefont{Pretorius}},
  \bibinfo{journal}{Class. Quant. Grav.} \textbf{\bibinfo{volume}{22}},
  \bibinfo{pages}{425} (\bibinfo{year}{2005}{\natexlab{a}}),
  \eprint{gr-qc/0407110}.

\bibitem[{\citenamefont{Brandt et~al.}(2000)\citenamefont{Brandt, Correll,
  G\'{o}mez, Huq, Laguna, Lehner, Marronetti, Matzner, Neilsen, Pullin
  et~al.}}]{Brandt00}
\bibinfo{author}{\bibfnamefont{S.}~\bibnamefont{Brandt}},
  \bibinfo{author}{\bibfnamefont{R.}~\bibnamefont{Correll}},
  \bibinfo{author}{\bibfnamefont{R.}~\bibnamefont{G\'{o}mez}},
  \bibinfo{author}{\bibfnamefont{M.~F.} \bibnamefont{Huq}},
  \bibinfo{author}{\bibfnamefont{P.}~\bibnamefont{Laguna}},
  \bibinfo{author}{\bibfnamefont{L.}~\bibnamefont{Lehner}},
  \bibinfo{author}{\bibfnamefont{P.}~\bibnamefont{Marronetti}},
  \bibinfo{author}{\bibfnamefont{R.~A.} \bibnamefont{Matzner}},
  \bibinfo{author}{\bibfnamefont{D.}~\bibnamefont{Neilsen}},
  \bibinfo{author}{\bibfnamefont{J.}~\bibnamefont{Pullin}},
  \bibnamefont{et~al.}, \bibinfo{journal}{Phys. Rev. Lett.}
  \textbf{\bibinfo{volume}{85}}, \bibinfo{pages}{5496} (\bibinfo{year}{2000}).

\bibitem[{\citenamefont{Alcubierre et~al.}(2004)\citenamefont{Alcubierre,
  Br\"ugmann, Diener, Guzm\'an, Hawke, Hawley, Herrmann, Koppitz, Pollney,
  Seidel et~al.}}]{Alcubierre2003:pre-ISCO-coalescence-times}
\bibinfo{author}{\bibfnamefont{M.}~\bibnamefont{Alcubierre}},
  \bibinfo{author}{\bibfnamefont{B.}~\bibnamefont{Br\"ugmann}},
  \bibinfo{author}{\bibfnamefont{P.}~\bibnamefont{Diener}},
  \bibinfo{author}{\bibfnamefont{F.~S.} \bibnamefont{Guzm\'an}},
  \bibinfo{author}{\bibfnamefont{I.}~\bibnamefont{Hawke}},
  \bibinfo{author}{\bibfnamefont{S.}~\bibnamefont{Hawley}},
  \bibinfo{author}{\bibfnamefont{F.}~\bibnamefont{Herrmann}},
  \bibinfo{author}{\bibfnamefont{M.}~\bibnamefont{Koppitz}},
  \bibinfo{author}{\bibfnamefont{D.}~\bibnamefont{Pollney}},
  \bibinfo{author}{\bibfnamefont{E.}~\bibnamefont{Seidel}},
  \bibnamefont{et~al.}, \bibinfo{journal}{Phys.Rev.D}
  \textbf{\bibinfo{volume}{72}}, \bibinfo{pages}{044004}
  (\bibinfo{year}{2004}), \eprint{gr-qc/0411149}.

\bibitem[{\citenamefont{Imbiriba et~al.}(2004)\citenamefont{Imbiriba, Baker,
  Choi, Centrella, Fiske, Brown, van Meter, and Olson}}]{Imbiriba:2004tp}
\bibinfo{author}{\bibfnamefont{B.}~\bibnamefont{Imbiriba}},
  \bibinfo{author}{\bibfnamefont{J.}~\bibnamefont{Baker}},
  \bibinfo{author}{\bibfnamefont{D.-I.} \bibnamefont{Choi}},
  \bibinfo{author}{\bibfnamefont{J.}~\bibnamefont{Centrella}},
  \bibinfo{author}{\bibfnamefont{D.~R.} \bibnamefont{Fiske}},
  \bibinfo{author}{\bibfnamefont{J.~D.} \bibnamefont{Brown}},
  \bibinfo{author}{\bibfnamefont{J.~R.} \bibnamefont{van Meter}},
  \bibnamefont{and} \bibinfo{author}{\bibfnamefont{K.}~\bibnamefont{Olson}},
  \bibinfo{journal}{Phys. Rev. D} \textbf{\bibinfo{volume}{70}},
  \bibinfo{pages}{124025} (\bibinfo{year}{2004}), \eprint{gr-qc/0403048}.

\bibitem[{\citenamefont{Fiske et~al.}(2005)\citenamefont{Fiske, Baker, van
  Meter, Choi, and Centrella}}]{Fiske:2005fx}
\bibinfo{author}{\bibfnamefont{D.~R.} \bibnamefont{Fiske}},
  \bibinfo{author}{\bibfnamefont{J.~G.} \bibnamefont{Baker}},
  \bibinfo{author}{\bibfnamefont{J.~R.} \bibnamefont{van Meter}},
  \bibinfo{author}{\bibfnamefont{D.-I.} \bibnamefont{Choi}}, \bibnamefont{and}
  \bibinfo{author}{\bibfnamefont{J.~M.} \bibnamefont{Centrella}},
  \bibinfo{journal}{Phys. Rev. D} \textbf{\bibinfo{volume}{71}},
  \bibinfo{pages}{104036} (\bibinfo{year}{2005}), \eprint{gr-qc/0503100}.

\bibitem[{\citenamefont{Sperhake et~al.}(2005)\citenamefont{Sperhake, Kelly,
  Laguna, Smith, and Schnetter}}]{Sperhake:2005uf}
\bibinfo{author}{\bibfnamefont{U.}~\bibnamefont{Sperhake}},
  \bibinfo{author}{\bibfnamefont{B.}~\bibnamefont{Kelly}},
  \bibinfo{author}{\bibfnamefont{P.}~\bibnamefont{Laguna}},
  \bibinfo{author}{\bibfnamefont{K.~L.} \bibnamefont{Smith}}, \bibnamefont{and}
  \bibinfo{author}{\bibfnamefont{E.}~\bibnamefont{Schnetter}},
  \bibinfo{journal}{Phys. Rev. D} \textbf{\bibinfo{volume}{71}},
  \bibinfo{pages}{124042} (\bibinfo{year}{2005}), \eprint{gr-qc/0503071}.

\bibitem[{\citenamefont{Schnetter et~al.}(2004)\citenamefont{Schnetter, Hawley,
  and Hawke}}]{Schnetter-etal-03b}
\bibinfo{author}{\bibfnamefont{E.}~\bibnamefont{Schnetter}},
  \bibinfo{author}{\bibfnamefont{S.~H.} \bibnamefont{Hawley}},
  \bibnamefont{and} \bibinfo{author}{\bibfnamefont{I.}~\bibnamefont{Hawke}},
  \bibinfo{journal}{Class. Quantum Grav.} \textbf{\bibinfo{volume}{21}},
  \bibinfo{pages}{1465} (\bibinfo{year}{2004}), \eprint{gr-qc/0310042}.

\bibitem[{\citenamefont{Pretorius and Lehner}(2004)}]{Pretorius:2003wc}
\bibinfo{author}{\bibfnamefont{F.}~\bibnamefont{Pretorius}} \bibnamefont{and}
  \bibinfo{author}{\bibfnamefont{L.}~\bibnamefont{Lehner}},
  \bibinfo{journal}{J. Comput. Phys.} \textbf{\bibinfo{volume}{198}},
  \bibinfo{pages}{10} (\bibinfo{year}{2004}), \eprint{gr-qc/0302003}.

\bibitem[{\citenamefont{Pretorius and Choptuik}(2005)}]{Pretorius:2005ua}
\bibinfo{author}{\bibfnamefont{F.}~\bibnamefont{Pretorius}} \bibnamefont{and}
  \bibinfo{author}{\bibfnamefont{M.~W.} \bibnamefont{Choptuik}}
  (\bibinfo{year}{2005}), \eprint{gr-qc/0508110}.

\bibitem[{\citenamefont{Br\"ugmann et~al.}(2004)\citenamefont{Br\"ugmann,
  Tichy, and Jansen}}]{Bruegmann:2003aw}
\bibinfo{author}{\bibfnamefont{B.}~\bibnamefont{Br\"ugmann}},
  \bibinfo{author}{\bibfnamefont{W.}~\bibnamefont{Tichy}}, \bibnamefont{and}
  \bibinfo{author}{\bibfnamefont{N.}~\bibnamefont{Jansen}},
  \bibinfo{journal}{Phys. Rev. Lett.} \textbf{\bibinfo{volume}{92}},
  \bibinfo{pages}{211101} (\bibinfo{year}{2004}), \eprint{gr-qc/0312112}.

\bibitem[{\citenamefont{Pretorius}(2005{\natexlab{b}})}]{Pretorius:2005gq}
\bibinfo{author}{\bibfnamefont{F.}~\bibnamefont{Pretorius}},
  \bibinfo{journal}{Phys. Rev. Lett.} \textbf{\bibinfo{volume}{95}},
  \bibinfo{pages}{121101} (\bibinfo{year}{2005}{\natexlab{b}}),
  \eprint{gr-qc/0507014}.

\bibitem[{\citenamefont{Zerilli}(1970)}]{Zerilli70}
\bibinfo{author}{\bibfnamefont{F.~J.} \bibnamefont{Zerilli}},
  \bibinfo{journal}{Phys. Rev. Lett.} \textbf{\bibinfo{volume}{24}},
  \bibinfo{pages}{737} (\bibinfo{year}{1970}).

\bibitem[{\citenamefont{Abrahams et~al.}(1998)\citenamefont{Abrahams, Rezzolla,
  Rupright, Anderson, Anninos, Baumgarte, Bishop, Brandt, Browne, Camarda
  et~al.}}]{Abrahams97a}
\bibinfo{author}{\bibfnamefont{A.~M.} \bibnamefont{Abrahams}},
  \bibinfo{author}{\bibfnamefont{L.}~\bibnamefont{Rezzolla}},
  \bibinfo{author}{\bibfnamefont{M.~E.} \bibnamefont{Rupright}},
  \bibinfo{author}{\bibfnamefont{A.}~\bibnamefont{Anderson}},
  \bibinfo{author}{\bibfnamefont{P.}~\bibnamefont{Anninos}},
  \bibinfo{author}{\bibfnamefont{T.~W.} \bibnamefont{Baumgarte}},
  \bibinfo{author}{\bibfnamefont{N.~T.} \bibnamefont{Bishop}},
  \bibinfo{author}{\bibfnamefont{S.~R.} \bibnamefont{Brandt}},
  \bibinfo{author}{\bibfnamefont{J.~C.} \bibnamefont{Browne}},
  \bibinfo{author}{\bibfnamefont{K.}~\bibnamefont{Camarda}},
  \bibnamefont{et~al.}, \bibinfo{journal}{Phys. Rev. Lett.}
  \textbf{\bibinfo{volume}{80}}, \bibinfo{pages}{1812} (\bibinfo{year}{1998}),
  \eprint{gr-qc/9709082}.

\bibitem[{\citenamefont{Beetle et~al.}(2005)\citenamefont{Beetle, Bruni, Burko,
  and Nerozzi}}]{Beetle:2004wu}
\bibinfo{author}{\bibfnamefont{C.}~\bibnamefont{Beetle}},
  \bibinfo{author}{\bibfnamefont{M.}~\bibnamefont{Bruni}},
  \bibinfo{author}{\bibfnamefont{L.~M.} \bibnamefont{Burko}}, \bibnamefont{and}
  \bibinfo{author}{\bibfnamefont{A.}~\bibnamefont{Nerozzi}},
  \bibinfo{journal}{Phys. Rev. D} \textbf{\bibinfo{volume}{72}},
  \bibinfo{pages}{024013} (\bibinfo{year}{2005}), \eprint{gr-qc/0407012}.

\bibitem[{\citenamefont{Burko et~al.}(2005)\citenamefont{Burko, Baumgarte, and
  Beetle}}]{Burko:2005fa}
\bibinfo{author}{\bibfnamefont{L.~M.} \bibnamefont{Burko}},
  \bibinfo{author}{\bibfnamefont{T.~W.} \bibnamefont{Baumgarte}},
  \bibnamefont{and} \bibinfo{author}{\bibfnamefont{C.}~\bibnamefont{Beetle}},
  \bibinfo{journal}{Phys. Rev. D} \textbf{\bibinfo{volume}{73}},
  \bibinfo{pages}{024002} (\bibinfo{year}{2005}), \eprint{gr-qc/0505028}.

\bibitem[{\citenamefont{Beetle and Burko}(2002)}]{Beetle:2002iu}
\bibinfo{author}{\bibfnamefont{C.}~\bibnamefont{Beetle}} \bibnamefont{and}
  \bibinfo{author}{\bibfnamefont{L.~M.} \bibnamefont{Burko}},
  \bibinfo{journal}{Phys. Rev. Lett.} \textbf{\bibinfo{volume}{89}},
  \bibinfo{pages}{271101} (\bibinfo{year}{2002}), \eprint{gr-qc/0210019}.

\bibitem[{\citenamefont{Nerozzi et~al.}(2005)\citenamefont{Nerozzi, Beetle,
  Bruni, Burko, and Pollney}}]{Nerozzi:2004wv}
\bibinfo{author}{\bibfnamefont{A.}~\bibnamefont{Nerozzi}},
  \bibinfo{author}{\bibfnamefont{C.}~\bibnamefont{Beetle}},
  \bibinfo{author}{\bibfnamefont{M.}~\bibnamefont{Bruni}},
  \bibinfo{author}{\bibfnamefont{L.~M.} \bibnamefont{Burko}}, \bibnamefont{and}
  \bibinfo{author}{\bibfnamefont{D.}~\bibnamefont{Pollney}},
  \bibinfo{journal}{Phys. Rev.} \textbf{\bibinfo{volume}{D72}},
  \bibinfo{pages}{024014} (\bibinfo{year}{2005}), \eprint{gr-qc/0407013}.

\bibitem[{\citenamefont{Nerozzi et~al.}(2006)\citenamefont{Nerozzi, Bruni, Re,
  and Burko}}]{Nerozzi:2005hz}
\bibinfo{author}{\bibfnamefont{A.}~\bibnamefont{Nerozzi}},
  \bibinfo{author}{\bibfnamefont{M.}~\bibnamefont{Bruni}},
  \bibinfo{author}{\bibfnamefont{V.}~\bibnamefont{Re}}, \bibnamefont{and}
  \bibinfo{author}{\bibfnamefont{L.~M.} \bibnamefont{Burko}},
  \bibinfo{journal}{Phys. Rev. D} \textbf{\bibinfo{volume}{73}},
  \bibinfo{pages}{044020} (\bibinfo{year}{2006}), \eprint{gr-qc/0507068}.

\bibitem[{\citenamefont{Krivan et~al.}(1997)\citenamefont{Krivan, Laguna,
  Papadopoulos, and Andersson}}]{Krivan97a}
\bibinfo{author}{\bibfnamefont{W.}~\bibnamefont{Krivan}},
  \bibinfo{author}{\bibfnamefont{P.}~\bibnamefont{Laguna}},
  \bibinfo{author}{\bibfnamefont{P.}~\bibnamefont{Papadopoulos}},
  \bibnamefont{and}
  \bibinfo{author}{\bibfnamefont{N.}~\bibnamefont{Andersson}},
  \bibinfo{journal}{Phys. Rev. D} \textbf{\bibinfo{volume}{56}},
  \bibinfo{pages}{3395} (\bibinfo{year}{1997}).

\bibitem[{\citenamefont{Pazos-\'Avalos and Lousto}(2004)}]{Pazos-Avalos:2004rp}
\bibinfo{author}{\bibfnamefont{E.}~\bibnamefont{Pazos-\'Avalos}}
  \bibnamefont{and} \bibinfo{author}{\bibfnamefont{C.~O.}
  \bibnamefont{Lousto}}, \bibinfo{journal}{Phys. Rev. D}
  \textbf{\bibinfo{volume}{72}}, \bibinfo{pages}{084022}
  (\bibinfo{year}{2004}), \eprint{gr-qc/0409065}.

\bibitem[{\citenamefont{Baker and Campanelli}(2000)}]{Baker00a}
\bibinfo{author}{\bibfnamefont{J.}~\bibnamefont{Baker}} \bibnamefont{and}
  \bibinfo{author}{\bibfnamefont{M.}~\bibnamefont{Campanelli}},
  \bibinfo{journal}{Phys. Rev. D} \textbf{\bibinfo{volume}{62}},
  \bibinfo{pages}{127501} (\bibinfo{year}{2000}).

\bibitem[{\citenamefont{{T}eukolsky}(1973)}]{Teukolsky73}
\bibinfo{author}{\bibfnamefont{S.~A.} \bibnamefont{{T}eukolsky}},
  \bibinfo{journal}{Astrophys. J.} \textbf{\bibinfo{volume}{185}},
  \bibinfo{pages}{635} (\bibinfo{year}{1973}).

\bibitem[{\citenamefont{Newman and Penrose}(1962)}]{Newman62a}
\bibinfo{author}{\bibfnamefont{E.}~\bibnamefont{Newman}} \bibnamefont{and}
  \bibinfo{author}{\bibfnamefont{R.}~\bibnamefont{Penrose}},
  \bibinfo{journal}{J. Math. Phys.} \textbf{\bibinfo{volume}{3}},
  \bibinfo{pages}{566} (\bibinfo{year}{1962}).

\bibitem[{\citenamefont{Szekeres}(1965)}]{Szekeres65}
\bibinfo{author}{\bibfnamefont{P.}~\bibnamefont{Szekeres}},
  \bibinfo{journal}{J. Math. Phys} \textbf{\bibinfo{volume}{6}},
  \bibinfo{pages}{1387} (\bibinfo{year}{1965}).

\bibitem[{\citenamefont{Chandrasekhar}(1983)}]{Chandrasekhar83}
\bibinfo{author}{\bibfnamefont{S.}~\bibnamefont{Chandrasekhar}},
  \emph{\bibinfo{title}{The Mathematical Theory of Black Holes}}
  (\bibinfo{publisher}{Oxford University Press}, \bibinfo{address}{Oxford,
  England}, \bibinfo{year}{1983}).

\bibitem[{\citenamefont{Kinnersley}(1969)}]{Kinnersley_1969}
\bibinfo{author}{\bibfnamefont{W.}~\bibnamefont{Kinnersley}},
  \bibinfo{journal}{J. Math. Phys.} \textbf{\bibinfo{volume}{10}},
  \bibinfo{pages}{1195} (\bibinfo{year}{1969}).

\bibitem[{\citenamefont{Smarr}(1977)}]{Smarr77}
\bibinfo{author}{\bibfnamefont{L.}~\bibnamefont{Smarr}}, \bibinfo{journal}{Ann.
  N. Y. Acad. Sci.} \textbf{\bibinfo{volume}{302}}, \bibinfo{pages}{569}
  (\bibinfo{year}{1977}).

\bibitem[{\citenamefont{Brandt}(1996)}]{Brandt96a}
\bibinfo{author}{\bibfnamefont{S.}~\bibnamefont{Brandt}}, Ph.D. thesis,
  \bibinfo{school}{University of Illinois at Urbana-Champaign},
  \bibinfo{address}{Urbana, Illinois} (\bibinfo{year}{1996}).

\bibitem[{\citenamefont{D.~Kramer and Herlt}(1980)}]{Kramer80}
\bibinfo{author}{\bibfnamefont{M.~M.} \bibnamefont{D.~Kramer},
  \bibfnamefont{H.~Stephani}} \bibnamefont{and}
  \bibinfo{author}{\bibfnamefont{E.}~\bibnamefont{Herlt}},
  \emph{\bibinfo{title}{Exact Solutions of {E}instein's Field Equations}}
  (\bibinfo{publisher}{Cambridge University Press},
  \bibinfo{address}{Cambridge}, \bibinfo{year}{1980}).

\bibitem[{\citenamefont{Matte}(1953)}]{Matte53}
\bibinfo{author}{\bibfnamefont{A.}~\bibnamefont{Matte}}, \bibinfo{journal}{Can.
  J. Math.} \textbf{\bibinfo{volume}{5}}, \bibinfo{pages}{1}
  (\bibinfo{year}{1953}).

\bibitem[{\citenamefont{Mars}(1999)}]{Mars_2001}
\bibinfo{author}{\bibfnamefont{M.}~\bibnamefont{Mars}},
  \emph{\bibinfo{title}{On the reconstruction of the {Kerr} background}}
  (\bibinfo{year}{1999}), \bibinfo{note}{(Unpublished notes)}.

\bibitem[{lap()}]{lapack}
\emph{\bibinfo{title}{Lapack: {\tt http://www.netlib.org/lapack/lug/}}}.

\bibitem[{\citenamefont{York}(1979)}]{York79}
\bibinfo{author}{\bibfnamefont{J.~W.} \bibnamefont{York}}, in
  \emph{\bibinfo{booktitle}{Sources of Gravitational Radiation}}, edited by
  \bibinfo{editor}{\bibfnamefont{L.~L.} \bibnamefont{Smarr}}
  (\bibinfo{publisher}{Cambridge University Press},
  \bibinfo{address}{Cambridge, UK}, \bibinfo{year}{1979}), pp.
  \bibinfo{pages}{83--126}, ISBN \bibinfo{isbn}{0-521-22778-X}.

\bibitem[{\citenamefont{Zlochower et~al.}(2005)\citenamefont{Zlochower, Baker,
  Campanelli, and Lousto}}]{Zlochower:2005bj}
\bibinfo{author}{\bibfnamefont{Y.}~\bibnamefont{Zlochower}},
  \bibinfo{author}{\bibfnamefont{J.~G.} \bibnamefont{Baker}},
  \bibinfo{author}{\bibfnamefont{M.}~\bibnamefont{Campanelli}},
  \bibnamefont{and} \bibinfo{author}{\bibfnamefont{C.~O.}
  \bibnamefont{Lousto}}, \bibinfo{journal}{Phys. Rev.}
  \textbf{\bibinfo{volume}{D72}}, \bibinfo{pages}{024021}
  (\bibinfo{year}{2005}), \eprint{gr-qc/0505055}.

\bibitem[{Cac()}]{Cactusweb}
\emph{\bibinfo{title}{Cactus: {\tt http://www.cactuscode.org}}}.

\bibitem[{\citenamefont{Thornburg}(2004)}]{Thornburg2003:AH-finding}
\bibinfo{author}{\bibfnamefont{J.}~\bibnamefont{Thornburg}},
  \bibinfo{journal}{Class. Quantum Grav.} \textbf{\bibinfo{volume}{21}},
  \bibinfo{pages}{743} (\bibinfo{year}{2004}), \eprint{gr-qc/0306056}.

\bibitem[{\citenamefont{Bowen and York}(1980)}]{Bowen80}
\bibinfo{author}{\bibfnamefont{J.~M.} \bibnamefont{Bowen}} \bibnamefont{and}
  \bibinfo{author}{\bibfnamefont{J.~W.} \bibnamefont{York},
  \bibfnamefont{Jr.}}, \bibinfo{journal}{Phys. Rev. D}
  \textbf{\bibinfo{volume}{21}}, \bibinfo{pages}{2047} (\bibinfo{year}{1980}).

\bibitem[{\citenamefont{Gleiser et~al.}(1998)\citenamefont{Gleiser, Nicasio,
  Price, and Pullin}}]{Gleiser:1998ng}
\bibinfo{author}{\bibfnamefont{R.~J.} \bibnamefont{Gleiser}},
  \bibinfo{author}{\bibfnamefont{C.~O.} \bibnamefont{Nicasio}},
  \bibinfo{author}{\bibfnamefont{R.~H.} \bibnamefont{Price}}, \bibnamefont{and}
  \bibinfo{author}{\bibfnamefont{J.}~\bibnamefont{Pullin}},
  \bibinfo{journal}{Phys. Rev. D} \textbf{\bibinfo{volume}{57}},
  \bibinfo{pages}{3401} (\bibinfo{year}{1998}), \eprint{gr-qc/9710096}.

\bibitem[{\citenamefont{Alcubierre et~al.}(2003)\citenamefont{Alcubierre,
  Br\"ugmann, Diener, Koppitz, Pollney, Seidel, and Takahashi}}]{Alcubierre02a}
\bibinfo{author}{\bibfnamefont{M.}~\bibnamefont{Alcubierre}},
  \bibinfo{author}{\bibfnamefont{B.}~\bibnamefont{Br\"ugmann}},
  \bibinfo{author}{\bibfnamefont{P.}~\bibnamefont{Diener}},
  \bibinfo{author}{\bibfnamefont{M.}~\bibnamefont{Koppitz}},
  \bibinfo{author}{\bibfnamefont{D.}~\bibnamefont{Pollney}},
  \bibinfo{author}{\bibfnamefont{E.}~\bibnamefont{Seidel}}, \bibnamefont{and}
  \bibinfo{author}{\bibfnamefont{R.}~\bibnamefont{Takahashi}},
  \bibinfo{journal}{Phys. Rev. D} \textbf{\bibinfo{volume}{67}},
  \bibinfo{pages}{084023} (\bibinfo{year}{2003}), \eprint{gr-qc/0206072}.

\bibitem[{\citenamefont{Dain et~al.}(2002)\citenamefont{Dain, Lousto, and
  Takahashi}}]{Dain:2002ee}
\bibinfo{author}{\bibfnamefont{S.}~\bibnamefont{Dain}},
  \bibinfo{author}{\bibfnamefont{C.~O.} \bibnamefont{Lousto}},
  \bibnamefont{and}
  \bibinfo{author}{\bibfnamefont{R.}~\bibnamefont{Takahashi}},
  \bibinfo{journal}{Phys. Rev. D} \textbf{\bibinfo{volume}{65}},
  \bibinfo{pages}{104038} (\bibinfo{year}{2002}),
  \eprint[http://arXiv.org/abs]{gr-qc/0201062}.

\bibitem[{\citenamefont{Abrahams and Price}(1996)}]{Abrahams95c}
\bibinfo{author}{\bibfnamefont{A.}~\bibnamefont{Abrahams}} \bibnamefont{and}
  \bibinfo{author}{\bibfnamefont{R.}~\bibnamefont{Price}},
  \bibinfo{journal}{Phys. Rev. D} \textbf{\bibinfo{volume}{53}},
  \bibinfo{pages}{1972} (\bibinfo{year}{1996}).

\bibitem[{\citenamefont{Diener}()}]{Diener_comm}
\bibinfo{author}{\bibfnamefont{P.}~\bibnamefont{Diener}},
  \bibinfo{note}{private communication.}

\bibitem[{\citenamefont{Khanna et~al.}(1999)\citenamefont{Khanna, Baker,
  Gleiser, Laguna, Nicasio, Nollert, Price, and Pullin}}]{Khanna99a}
\bibinfo{author}{\bibfnamefont{G.}~\bibnamefont{Khanna}},
  \bibinfo{author}{\bibfnamefont{J.}~\bibnamefont{Baker}},
  \bibinfo{author}{\bibfnamefont{R.}~\bibnamefont{Gleiser}},
  \bibinfo{author}{\bibfnamefont{P.}~\bibnamefont{Laguna}},
  \bibinfo{author}{\bibfnamefont{C.}~\bibnamefont{Nicasio}},
  \bibinfo{author}{\bibfnamefont{H.-P.} \bibnamefont{Nollert}},
  \bibinfo{author}{\bibfnamefont{R.}~\bibnamefont{Price}}, \bibnamefont{and}
  \bibinfo{author}{\bibfnamefont{J.}~\bibnamefont{Pullin}},
  \bibinfo{journal}{Phys. Rev. Lett.} \textbf{\bibinfo{volume}{83}},
  \bibinfo{pages}{3581} (\bibinfo{year}{1999}).

\bibitem[{\citenamefont{Baumgarte}(2000)}]{Baumgarte00a}
\bibinfo{author}{\bibfnamefont{T.~W.} \bibnamefont{Baumgarte}},
  \bibinfo{journal}{Phys. Rev. D} \textbf{\bibinfo{volume}{62}},
  \bibinfo{pages}{024018} (\bibinfo{year}{2000}), \eprint{gr-qc/0004050}.

\bibitem[{\citenamefont{Cook}(1994)}]{Cook94}
\bibinfo{author}{\bibfnamefont{G.~B.} \bibnamefont{Cook}},
  \bibinfo{journal}{Phys. Rev. D} \textbf{\bibinfo{volume}{50}},
  \bibinfo{pages}{5025} (\bibinfo{year}{1994}).

\bibitem[{\citenamefont{Campanelli et~al.}(2005)\citenamefont{Campanelli,
  Lousto, Marronetti, and Zlochower}}]{Campanelli:2005dd}
\bibinfo{author}{\bibfnamefont{M.}~\bibnamefont{Campanelli}},
  \bibinfo{author}{\bibfnamefont{C.~O.} \bibnamefont{Lousto}},
  \bibinfo{author}{\bibfnamefont{P.}~\bibnamefont{Marronetti}},
  \bibnamefont{and} \bibinfo{author}{\bibfnamefont{Y.}~\bibnamefont{Zlochower}}
  (\bibinfo{year}{2005}), \eprint{gr-qc/0511048}.

\bibitem[{\citenamefont{Reimann and Br{\"u}gmann}(2004)}]{Reimann:2003zd}
\bibinfo{author}{\bibfnamefont{B.}~\bibnamefont{Reimann}} \bibnamefont{and}
  \bibinfo{author}{\bibfnamefont{B.}~\bibnamefont{Br{\"u}gmann}},
  \bibinfo{journal}{Phys. Rev. D} \textbf{\bibinfo{volume}{69}},
  \bibinfo{pages}{044006} (\bibinfo{year}{2004}), \eprint{gr-qc/0307036}.

\end{thebibliography}

\end{document}